\documentclass[pdflatex,sn-aps]{sn-jnl}% American Physical Society (APS) Reference Style
\usepackage{graphicx}
\usepackage{xcolor}
\usepackage{dsfont}
\DeclareFontFamily{OT1}{pzc}{}
\DeclareFontShape{OT1}{pzc}{m}{it}{<-> s * [1.100] pzcmi7t}{}
\DeclareMathAlphabet{\mathpzc}{OT1}{pzc}{m}{it}
\DeclareMathAlphabet{\mathpzc}{OT1}{pzc}{m}{it}
\DeclareMathOperator\arctanh{arctanh}

\usepackage{textcomp}
\usepackage{mathtools}
\DeclarePairedDelimiter\bra{\langle}{\rvert}
\DeclarePairedDelimiter\ket{\lvert}{\rangle}
\DeclarePairedDelimiterX\braket[2]{\langle}{\rangle}{#1 \delimsize\vert #2}
\jyear{2022}
\theoremstyle{thmstyleone}

\theoremstyle{thmstyletwo}

\theoremstyle{thmstylethree}

\begin{document}
\title[Conformal Dilaton-Higgs Gravity on Warped Spacetimes]{Conformal Dilaton-Higgs Gravity on Warped Spacetimes: Black Hole Paradoxes Revisited}
\author*[1,2]{\fnm{Reinoud Jan} \sur{Slagter}}\email{info@asfyon.com\footnote{or: reinoudslagter@gmail.com}}
\affil*[1]{\orgdiv{Asfyon}, \orgname{Astronomisch Fysisch Onderzoek Nederland} \orgaddress{\street{}, \city{Bussum}, \postcode{} \state{} \country{The Netherlands}}}
\affil[2]{\orgdiv{former:} \orgname{University of Amsterdam}, \orgaddress{\street{} \city{} \postcode{Department of Theoretical Physics}, \state{} \country{The Netherlands}}}
%=========================================================================================
\abstract{We investigate on a Randall-Sundrum warped spacetime, a Kerr-like black hole  in the conformal dilaton-Higgs $(\omega,\Phi)$ gravity model.
We applied the antipodal boundary condition on the Klein surface using the $\mathds{Z}_2$-symmetry in the "large" (bulk) extra dimension. It turns out that the pseudo-Riemannian 5D manifold can be written as an effective 4D Riemannian brane  spacetime, $\mathds{R}^2_+\times\mathds{R}^1\times S^1$, where $\mathds{R}^2_+$ is conformally flat. The solution in valid on both manifolds. So the solution can equally well  described by an instanton solution. An advantage is that  antipodicity can be maintained without a "cut-and-past" method or to rely on quantum cloning, when treating  the scattering description of the evaporation process of the Hawking radiation. We need the windingnumber as quantum number.
Moreover, the equations are invariant under time reversal.
The problem of finding the matching condition of the near-horizon approximation and the far-away Regge-Wheeler approximation, can possibly be solved by splitting the spacetime in a dilaton field times an "un-physical" spacetime, which is conformally flat.
In the case of a constant gauge field, we find that the conform invariant mass term $\sim \Phi^2\omega^2$ in the Lagrangian follows directly from the superfluous dilaton equation by suitable choice of the scale of the extra dimension. 
Finally, we bring forward the relation between the embedded Klein surface in $\mathds{R}^4$ and the quantum mechanical information paradox. }

\keywords{Conformal invariance, Dilaton-Higgs fields, Warped spacetimes, Antipodal boundary condition, Klein surface, Instanton, Black hole paradoxes, Minimal surfaces }
\maketitle
%=========================================================================================
\section{Introduction}\label{sec1}
%===========================================================================================
One of the most profound unsolved problems in theoretical physics is the discrepancy between quantum mechanics (QM) and general relativity theory (GRT).
One should like to obtain a consistent theory of quantum gravity, a major goal of theoretical physics (the "holy grail") that would reconcile QM and Einstein's GRT.
It is conjectured that the problem  how to handle quantum-gravity effects, will be found  near the horizon of black holes.

Back holes are in quite another perspective, incredibly unique and fascinating objects.
The black hole can take part in the weird and  wonderful behavior of QM. Particles can be in the quantum superposition of multiple states. They will be entangled until  observers will act on the system (¨collapse of the wavefunction¨). So new Planck scale physics should be necessary to address the information paradox, complementarity and to avoid the firewall.

During the last decades, several authors proposed different approaches in order to attack this epic problem. 
Pioneer work was delivered by Hawking\cite{hawking1975}, who investigated particle creation near the horizon of a  black hole. He found that black holes radiate as black bodies with a thermal spectrum. So the black hole would be in a mixed state.
However, this seems to suggest  violation of unitary evolution, because all information connected with the in-falling particle (¨in-state¨), appears to have lost after the evaporation of the black hole. In quantum field theory (QFT) one usually deals with pure quantum states which evolve unitarily. 
A possible solution was addressed by Almheiri, et al.,\cite{alm2013}, in order to solve the conflict between complementarity and firewalls. The in-falling observer would burn up at the horizon´s firewall. However, this viewpoint conflicts GRT.
Maldacena, et al,\cite{maldacena2013} proposed the ¨ER = EPR¨ hypothesis, which is based on the suggestion that connectivity in spacetime is equivalent to quantum entanglement. They assert that quantum entanglement of distant black holes is equivalent to the existence of an Einstein-Rosen bridge connecting them. However, one ignores the gravitational interaction between the two regions  I and II in the Penrose diagram.
Also the emergence of string theory, holography and gauge-gravity duality could shed light on this problem.
However, the information retrieval process is not quite clear\cite{mathur2009}. 
Recent investigation by Foo, et al.\cite{foo2021}, suggests that even so-called “spacetime superpositions” could exist, i.e., that quantum superpositions of different spacetimes not related by a global coordinate transformation could be possible\footnote{after an proposal of Bekenstein long time ago\cite{bekenstein1974}.}.

In general, one can state that in all these models the dynamical back reaction is ignored.
For example, the evaporation process will have an impact on the near-horizon spacetime. There will be a strong interaction between, for example,  a scalar field and the gravitation degrees of freedom\cite{betz2021}.

The departure from Hawking`s calculations, leads to quantum mechanical scattering algebra, realized by in- and out-going wave functions near the horizon\cite{thooft2015}.
Later, it was argued that one can reformulate this departure, by considering soft graviton exchange between matter fields near the horizon\footnote{See the clear overview of Betzios, et al.\cite{betz2021}.}.
Away from the horizon, one can then introduce an effective Regge-Wheeler gravitational potential. However, in all these models, the calculations on the quasi-normal modes were done on a fixed Schwarzschild background geometry, which is nor adequate for black hole evaporation process: the potential is time dependent.
One should like to obtain a complete scattering description for the outside observer, which is consistent with the local observer\footnote{Remember, an freely in-falling observer perceive spacetime as Minkowski, so will not notice the horizon.}.

This complementarity issue, which means the discrepancy between the local observer and the far away observer, can be reformulated in conformal dilaton gravity (CDG) by introducing a dilaton field $\omega$ and by writing $g_{\mu\nu}=\omega^{\frac{4}{d-2}}\tilde g_{\mu\nu}$\cite{thooft2015c}. Different observers experience different notions of $\omega$, i.e., the scale of spacetime. Moreover, $\omega$ (and a scalar field $\Phi$) and $\tilde g_{\mu\nu}$ are invariant under $\tilde g_{\mu\nu}\rightarrow \Omega^{\frac{4}{d-2}} \tilde g_{\mu\nu}, \omega \rightarrow \Omega^{-\frac{d-2}{2}}\omega, \Phi \rightarrow \Phi^{-\frac{d-2}{2}}\omega$. $\Omega$ can be used, for example, to make Ricci scalar for the local observer zero. 
Further, one observer sees Hawking radiation as matter, while on other one as part of his vacuum. The in-falling observer  will not notice any change in the mass of the black hole. The out side observer, on the other hand, observes a gradually shrinking mass.
It is conjectured that this conformal description will be necessary in order understand  the black hole paradoxes.

The approach we will address in this manuscript, relies on the antipodal boundary condition on a warped 5D Spacetime. This idea  was introduces decades ago by Schr\"odinger\cite{schrod1957}. It is sometimes called the elliptic interpretation of spacetime. 
Interesting investigation on this mapping where done by Sanchez, et al.\cite{san1987,san1988,fol1987} on a AdS spacetime.
The model was intensively extended by´t Hooft\cite{thooft2016,thooft2018b,thooft2019} in a slightly different setting, by considering a new treatment of the gravitational back-reaction. 
In brief, antipodal points on the horizon represent the same world-point or event. The spacetime inside  the horizon is removed such that the edges are glued together  by identifying the antipodes.
In the language of a Penrose diagram, "region II" is the antipode of "region I".

In order to avoid  wormhole constructions or demanding  "an other universe" in the Penrose diagram, it is essential that the asymptotic domain of the manifold maps one-to-one onto the ordinary spacetime in order to preserve the metric. 
In fact, one deals with one black hole.
´t Hooft performed calculations on the unitary evolution matrix   by using the  antipodal boundary condition, i.e., the transverse spherical coordinates $(\theta,\varphi)$  at region II, represent the antipode of region I. So there are no fixed points. In fact, there is no "hidden" sector in the Penrose diagram. The resulting Hartle-Hawking vacuum state  remains a pure state in stead of a thermodynamically mixed state for the outside observer. 
The Hawking particles are emerging from I, are maximally entangled  with particles emerging from II.
This approach has the potential to solve the information paradox and a firewall is not necessary.  The antipodal boundary condition\footnote{Also called the "cut-and-past" procedure or firewall transformation.},  minimizes the number of unconventional assumptions or to rely on full string theory and ¨fuzzyballs". It will require that the local laws of physics are invariant across the black hole boundary\cite{thooft2021}. 
Usually, one considers the Hawking particles as excitations of low energy particles that fulfill the standard model and applies perturbative quantum gravity on a fixed background metric (i.e., the gravitons). The distant observer notices the Hilbert space of low energy particles at given time. One then maps  the states at later times,  1 on 1 on earlier times.
One basically bypasses  the Planck-area. It manifests itself only by the higher modes of the spherical harmonics (in the angle variables). 
An unitary S matrix can then be constructed if one applies a cut-off for the higher modes of the harmonics.
In a recent study,´t Hooft\cite{thooft2022} tries to avoid the antipodal boundary condition, in order to overcome some inconsistencies, into a quantum clone description, in spherical symmetric case. 

In our 5D warped spacetime, however, this change is not necessary, due to the cylindrical symmetry of the Kerr-type spacetime we considered. It is strongly conjectured that the center of galaxies harbors a spinning Kerr black holes and not a Schwarzschild black hole. So there is a preferred direction, i.e., the $z$-axis. If one omits the $dz^2$, one obtains the well known exact  Ba\u nados-Teitelboim-Zanelli (BTZ) black hole solution in $(2+1)$-dimensional spacetime\cite{banadoz1992,slagter2019b}.

In the recently found  vacuum solution\cite{slagter2022c}, no further inconsistency was encountered, because  we had only to deal with the gravitational  freedom, i.e., the  dilaton field $\omega$, which  was solved exactly\footnote{Note that, due to  Birkhoff's theorem, spherical symmetric objects will not emit gravitational radiation.}.

In the present case, the scalar field will be written as $\Phi=\eta X(t,r)e^{in\varphi}Y_m(\varphi)$ and with the gauge field $A \sim \frac{n}{e}$.
In order find a consistent description of the antipodal map, the warped fifth dimension is mandatory,  in order to apply the non-orientable Klein surface.

In our model we will also try  to avoid in the dynamical evolution of the fields (scalar, dilaton and metric components) on the Kerr spacetime, the late-times and early-times behavior, by writing the PDE´s in Kruskal coordinates\footnote{We will work in polar coordinates, because the Kerr spacetime is axisymmetric. In general, the  field variables will then be written as $F(t,r,z,\varphi, y)=F(U,V)H(z,\varphi)K(y)$, with  $H$ cylindrical harmonics, $(U,V)$ Kruskal coordinates and $y$ the RS bulk dimension. Note: the axisymmetry is also needed when the scalar-gauge field, considered as vortex, is incorporated.}

We know that local QFT and GRT are invariant under  CPT transformations. In the model we consider here, we advocate that  antipodal map preserves also CPT. The basic argument being that CPT is an exact symmetry of nature, i. e., of QFT around flat background. It has the potential to be elevated to a form of discrete gauge symmetry in a full theory of quantum gravity.
One says that region II is a CPT-transformed quantum copy of region I.

There are some other approaches to this invariance. The conformal group implies Poincar\'e invariance and so PT invariance\cite{mannheim2013}. The PT symmetry would be a necessary and sufficient condition for unitary time evolution whereas hermiticity is only a sufficient condition. 
Dropping the hermiticity in favor of PT symmetry is not generally accepted
(see also the discussion by Betzios, et al.\cite{bet2016,bet2017} and references therein). 

There are several  reasons for considering a warped spacetime of Randall and Sundrum (RS)\cite{ran1999a,ran1999b}.
The recently found exact solution of the Kerr-like spacetime without a scalar field, was found in the CDG model on a warped RS spacetime\cite{slagter2018,slagter2021,slagter2022b,slagter2022c}.
These so-called  brane-world models provide a simplification of the full string model.
In the later string models, the extra dimensions are compactified or fold in on themselves in many ways, meaning that there are to many possible solutions to be able to make a clear prediction. 
In the RS model, only gravity can propagate into the bulk, while all other fields resides on the brane. 
Einstein gravity on the brane will be modified by the very embedding  itself and opens up a possible new way to address the  dark energy problem\cite{mann2005}. In a former study\cite{slagterpan2016} we applied this model on a Friedmann-Lema\^itre-Robertson-alker (FLRW) spacetime.  There is a contribution from the projected 5D Weyl tensor on the effective 4D brane, which carries information of the gravitational field outside the brane.
If one writes the 5D Einstein equation in CDG setting, it could be possible that  an effective theory can be constructed  without an UV cutoff, because the  fundamental scale $M_{5}$ can be much less than the effective scale $M_{Pl}$ due to the warp factor. The physical scale is therefore not determined by $M_{Pl}$.
There are some other arguments which advocate for the 5D model. 
First, the warped model could possibly solve the hierarchy problem.
Secondly, the description of the  antipodal boundary condition by means of the M\"obius strip in the 4D model, can be extended by considering the Klein surface, which can be embedded in $\mathds{R}^4$.
Quite recently, Maldacena, et al.\cite{maldacena2021}, found an interesting wormhole solution on the RS model. 
It is remarkable that the found exact  spacetime solution $\tilde g_{\mu\nu}$ together with the dilaton solution, is the same for the 5D and effective 4D Einstein equations. In a covariant approach of the model, it is mandatory to solve these equations together in order to obtain    consistency\cite{shirom2000,shirom2003}. Moreover, it turns out that the solution can be maintained, when one switches to the Riemannian 5D spacetime (Wick rotation).
This instanton solution can then be embedded in $\mathds{R}^4$, i.e., our effective brane in the RS model. This becomes clear when one complexifies the space as $\mathds{C}^1\times\mathds{C}^1$. By the Hopf fibration, one then makes the connection with our real space.

As already mentioned, the BTZ solution shares some features with the 4D Kerr solution and  gained new interest because it can be used to study quantum gravity issues. The model needs a cosmological constant and is asymptotically AdS.
The relation with the dynamical "uplifted"" BTZ solution\footnote{In this case the $dz^2$ term is maintained. It is remarkable that $\Lambda$ must then be taken zero.} was presented by Slagter\cite{slagter2019b}. 
After the discovery of the  AdS/CFT correspondence, the BTZ solution gained new interest and became a tool to understand black hole entropy. It is not yet clear whether pure 3D Einstein gravity make sense quantum mechanically without string theory embedding.
Adding a scalar field will make the model more realistic. The improvement of the treatment of the gravitational interaction of the propagating modes on a Schwarzschild background was done by Betzios, et al.\cite{betz2021}. They distinguish two regions. Far from the horizon the propagating fields evolve semi-classical under the Regge-Wheeler potential. In the near-horizon region, where the Regge-Wheeler potential is taken zero, they apply a gravitational scattering that captures non-perturbative  soft graviton exchange. An appropriate matching of the two dynamical systems is then necessary in order to obtain a unitary scattering matrix.

It is worth making some remarks on the recent work of Gaddam, et. al.\cite{gaddam2020,gaddam2021,gaddam2022,gaddam2022b}, on the
information paradox. They calculated the soft graviton exchange between the in- and out-going quanta, using the eikonal phase approach, in order to find the gravitation back reaction near the horizon. 
It is possible that the so-called multiple-scale wave approximation could be better applied\cite{choq1969,slagter2000}, in order to keep track of the different orders of approximation. The expansion parameter will then be the ratio of the background scale and the fast varying scale.
The expansion in harmonics will follow directly from the several orders of perturbation equations.

In this manuscript, we try to make no distinction between the two regions. By considering the conformal invariant model with the dilaton and scalar field, we  obtain two (quantum) interacting fields, which differ only in the effective potential. They contain also the metric component,  which determines the near horizon behavior. In the vacuum situation, an exact time dependent solution was found. In the new model, one can be guided by this solution, in order to obtain a numerical solution. Obstructions encountered in the antipodal boundary condition can be avoided in the 5D warped version: the antipodal map can be constructed on a Klein surface\cite{slagter2022c}.
It is interesting to compare the  dynamical behavior of the scale function $\omega$\footnote{which is, after all, part of the spacetime.} with the theory of gravitational echoes and so-called scrambling time\cite{abedi2017}.

The outline of this manuscript is the following. In section 2 we describe the new CDH model and the related, earlier found, vacuum solution. 
In section 3 we present the connection of the Klein surface with antipodicity.  In section 4 we outline some topological aspects. In section 5 we return to the information paradox in context with the Klein surface and in section 6 we briefly summarize the application of the high-frequency approximation.
%===============================================================================================
\section{The model}\label{sec2}
%============================================================================================
It is conjectured that a conformal invariant theory of gravity promises interesting results when applied to the quantum-gravity area. In particular, the conformal dilaton gravity model (CDG)\cite{alvarez2014, codello2013,thooft2015b,oda2015}.
We shall see that it is a route to construct a topologically regular theory of gravity.

A theory is called conformally invariant at the classical level, if its action is invariant under the conformal group of translations, dilatations LT and special conformal transformations\cite{felsager1998}. This is a local symmetry, if the metric is dynamically, as  will be considered here\footnote{It is a global symmetry, when one considers the spacetime fixed.}.
Let us consider the conformal invariant Lagrangian
\begin{eqnarray}
S=\int d^dx\sqrt{-\tilde g}\Bigl[-\frac{1}{2}\xi (\Phi\Phi^*+\omega^2)\tilde R-\frac{1}{2}\tilde g^{\mu\nu} \Bigl({\cal D}_\mu\Phi({\cal D}_\mu\Phi)^*+\partial_\mu\omega\partial_\nu\omega\Bigr)\cr
-\frac{1}{4}F_{\alpha\beta}F^{\mu\nu}-V(\Phi,\omega)-\Lambda\kappa^{\frac{4}{d-2}}\xi^{\frac{d}{d-2}}\omega^{\frac{2d}{d-2}} \Bigr],\label{2-1}
\end{eqnarray}
which is invariant under
\begin{equation}
\tilde g_{\mu\nu}\rightarrow \Omega^{\frac{4}{d-2}} \tilde g_{\mu\nu},\quad \omega \rightarrow \Omega^{-\frac{d-2}{2}}\omega, \quad \Phi \rightarrow \Omega^{-\frac{d-2}{2}}\Phi\label{2-2}
\end{equation}
for suitable choice of $V$ (for example the simplest one, $V=0$).
Further, $\xi=(d-2)/4(d-1)$.
The covariant derivative is taken with respect to the "un-physical" $\tilde g_{\mu\nu}$, which is defined in the CDG model, initiated by 't Hooft\cite{thooft2015b}, by writing
\begin{equation}
g_{\mu\nu}=\omega^{4/(d-2)} \tilde g_{\mu\nu}.\label{2-3}
\end{equation}
$\omega$ represents a dilaton field, which represents the scale dependency.
The gauge-covariant derivative is ${\cal D}_\mu\Phi = \tilde\nabla_\mu\Phi+ie A_\mu\Phi$ and  $F_{\mu\nu}$ the Abelian field strength. 
Further, we redefined the dilaton, $\omega^2\rightarrow -\frac{6\omega^2}{\kappa^2}$, in order to ensure that the field $\omega$ has the same unitarity and positivity properties as the scalar field.
The potential $V(\Phi ,\omega)$ is still unspecified. A massive term in $V(\Phi ,\omega)$ will break, in general, the tracelessness of the energy momentum tensor and therefore breaks the conformal invariance.
One considers then the symmetry as exact and spontaneously  broken, just as the BEH mechanism.
$\omega\rightarrow 0$ describes the small distance limit. No singularity should occur in this limit.
We parameterize the scalar field and gauge field as 
\begin{equation}
A_\mu=\Bigl[0,0,0,\frac{1}{e}(P(t,r)-n)\Bigr],\quad \Phi=\eta X(t,r) e^{in\varphi}.\label{2-4}
\end{equation}
The metric $\tilde g_{\mu\nu}$ is actually a metatensor. All the scale dependencies are contained in the dilaton and will be handled on equal footing with the scalar field and can be extended to scales close to the Planck scale.
We consider here the Kerr-like spacetime on a warped 5D spacetime with $\mathds{Z}_2$-symmetry \cite{slagter2021,slagter2022c}
\begin{equation}
ds^2=\omega(t,r,y)^{4/3}\Bigl[-N(t,r)^2dt^2+\frac{1}{N(t,r)^2}dr^2+dz^2+r^2(d\varphi+N^\varphi(t,r)dt)^2+d\mathpzc{y}^2\Bigr],\label{2-5}
\end{equation}
where $\mathpzc{y}$ is the extra dimension (not to confuse with the Cartesian y\footnote{One can also use the Eddington-Finkelstein coordinates $(U,r,z,\varphi,\mathds{y})$.}). Here $\omega$ was called a "warp factor" in the formulation of RS 5D warped spacetime with one large extra dimension and negative  bulk tension.  
It turns out that one can write   $\omega(t,r,\mathpzc{y})=\omega_1(t,r)\omega_2(\mathpzc{y})$, with $\omega_2(\mathpzc{y})=y_0$=constant (the length scale of the extra dimension).
The Standard Model (SM) fields are confined to the  4D brane, while gravity acts also in the fifth dimension. 
It possesses $\mathds{Z}_2$-symmetry, which means that when one approaches the brane from one side and go through it, one emerges into the bulk that looks the same, but with the normal reversed.
Since the pioneering publication of RS, many investigations were done in several related  domains.
In particular, Shiromizu et.al. \cite{shirom2003}, extended the RS model to a fully  covariant curvature formalism. See also the work of Maartens \cite{maartens2010}.
It this extended model, an  effective Einstein equation is found on the brane, with on the right hand side a contribution from the  5D Weyl tensor which carries information of the gravitational field  outside the brane. 
So the brane world observer may be subject to influences from the bulk.

One then obtains  field equations for the 5D spacetime together with the effective 4D Einstein equations (we took an empty bulk) \cite{slagter2021,slagter2022c}
\begin{equation}
{^{(5)}}{G_{\mu\nu}}=-\Lambda_5{^{(5)}g_{\mu\nu}},\label{2-6}
\end{equation}
\begin{equation}
{^{(4)}G_{\mu\nu}}=-\Lambda_{eff}{^{(4)}g_{\mu\nu}}+\kappa_4^2{^{(4)}T_{\mu\nu}}+\kappa_5^4{\cal S}_{\mu\nu}-{\cal E}_{\mu\nu},\label{2-7}
\end{equation}
where we have written
\begin{equation}
{^{(5)}g_{\mu\nu}}={^{(4)}g_{\mu\nu}}+n_\mu n_\nu,\label{2-8}
\end{equation}
with $n^\mu=[0,0,0,0,\frac{1}{\sqrt{y_0}}]$ the unit normal to the brane. Here ${^{(4)}T_{\mu\nu}}$ is the energy-momentum tensor on the brane, which also contains the contribution from the dilaton
\begin{equation}
{^{(4)}T_{\mu\nu}}=T^{(\omega)}_{\mu\nu}+T^{(scalar)}_{\mu\nu}\label{2-9}
\end{equation}
and ${\cal S}_{\mu\nu}$ the quadratic contribution of the energy-momentum tensor ${^{(4)}T_{\mu\nu}}$, arising from the extrinsic curvature terms in the projected Einstein tensor.
Further,
\begin{equation}
{\cal E}_{\mu\nu}={^{(5)}}C^\alpha_{\beta\rho\sigma}n_\alpha n^\rho {^{(4)}}g_{\mu}^{\beta}{^{(4)}}g_{\nu}^{\sigma},\label{2-10}
\end{equation}
represents the projection of the bulk Weyl tensor orthogonal to $n^\mu$. The effective gravitational field equations on the brane are not closed. One must solve at the same time the 5D gravitational field in the bulk.
For an empty bulk, there is no exchange of energy-momentum between bulk and brane. The interaction is purely gravitational. From the 4D conservation equations (contracted Bianchi identities), one then obtains $\nabla^\mu{\cal E}_{\mu\nu}=\kappa_5^4{\cal S}_{\mu\nu}$. It tells us that (1+3)-spacetime  variations in the matter-radiation can source KK modes.  
Now we replace again
\begin{equation}
{^{(4)}\tilde g_{\mu\nu}}\rightarrow \bar\omega^2 {^{(4)}\tilde g_{\mu\nu}}.\label{2-11}
\end{equation}
Variation of  the action Eq.(\ref{2-1}) for a general $V$\footnote{We omit for the time being the cosmological constant term $\sim\Lambda\omega^4$.} with respect to $\tilde g_{\mu\nu}$, $\Psi$ and $\omega$ yields ($d=4,5$)
\begin{eqnarray}
\xi\bar\omega\tilde R-\tilde g^{\mu\nu}\tilde\nabla_\mu\tilde\nabla_\nu\bar\omega-\frac{\partial V}{\partial\bar\omega}=0,\label{2-12}
\end{eqnarray}
\begin{eqnarray}
\xi\Phi\tilde R-\tilde g^{\mu\nu}{\cal D}_\mu{\cal D}_\nu\Phi-\frac{\partial V}{\partial\Phi^*}=0,\label{2-13}
\end{eqnarray}
and
\begin{eqnarray}
(\bar\omega^2+\Phi\Phi^*)\tilde G_{\mu\nu}=T^{(\Phi)}_{\mu\nu}+T_{\mu\nu}^{(\omega)}-\tilde g_{\mu\nu}V
-(\bar\omega^2+\Phi\Phi^*){\cal E}_{\mu\nu}\delta^4_d,\label{2-14}
\end{eqnarray}
with
\begin{eqnarray}
T_{\mu\nu}^{(\omega)}=\tilde\nabla_\mu\tilde\nabla_\nu\bar\omega^2-\tilde g_{\mu\nu}\tilde\nabla^2\bar\omega^2+\frac{1}{\xi}\Bigl(\frac{1}{2}\tilde g_{\alpha\beta}\tilde g_{\mu\nu}-\tilde g_{\mu\alpha}\tilde g_{\nu\beta}\Bigr)\partial^\alpha\bar\omega\partial^\beta\bar\omega,\label{2-15}
\end{eqnarray}
and 
\begin{equation}
T^{(\Phi)}_{\mu\nu}={\cal D}_\mu{\cal D}_\nu\Phi\Phi^*-\tilde g_{\mu\nu}{\cal D}^\alpha{\cal D}_\alpha\Phi\Phi^*+\frac{1}{\xi}\Bigl(\frac{1}{2}\tilde g_{\alpha\beta}\tilde g_{\mu\nu}-\tilde g_{\mu\alpha}\tilde g_{\nu\beta}\Bigr){\cal D}^\alpha\Phi{\cal D}^\beta\Phi^*\label{2-16}.
\end{equation}
Note that on the right hand side of the  4D effective Einstein equations now appears for $d=4$, the contribution from the bulk in the correct form. The only unknown functions are the metric components, the dilaton and scalar fields. $\omega$ and $\Phi$ can be treated on equal footing as  normal quantum fields on the small scale.
The dilaton equation Eq.(\ref{2-12}) is superfluous in the vacuum case\cite{slagter2022c}. In the general case with a scalar field, there will be an interaction between $\omega$ and $\Phi$ and some constraint on $V$.
%=========================================================================================
\subsection{The dilaton-Higgs model in conformal invariant  gravity}
%========================================================================================
\subsubsection{case:  $A_\mu(t,r)=0$}
%=======================================================================================
Firstly, we can switch off the  gauge potential, $A_\mu =0$.
From the Einstein field equations and the scalar equation one can isolate the PDE's for $N$, $\omega$ and $\Phi$:
\begin{eqnarray}
\ddot N=-N^4\Bigl(N''+3\frac{N'}{r}+\frac{N'^2}{N}\Bigr)+3\frac{\dot N^2}{N}\cr
\frac{1}{(\eta^2X^2+\omega^2)}\Bigl[6N^5(\eta^2X'^2+\omega^2)-6(\eta^2\dot X^2+\dot\omega^2)-\frac{3}{2r}N^5(\eta^2XX'+\omega\omega')\cr
+\frac{9}{2r^2}\eta^2n^2N^2X^2-6N^3\Bigl(\eta X\frac{\partial V}{\partial X}+\omega\frac{\partial V}{\partial\omega}+V \Bigr)\Bigr]\label{2-18}
\end{eqnarray}
\begin{eqnarray}
\ddot X=N^4\Bigl(X''+\frac{X'}{r}+2\frac{X'N'}{N}\Bigr)+2\frac{\dot X\dot N}{N}+\frac{2X}{(\eta^2 X^2+\omega^2)}\Bigl[N^4(\eta^2 X'^2+\omega'^2)\cr
-(\eta^2\dot X^2+\dot\omega^2)+N^2(V-\omega\frac{\partial V}{\partial\omega}) -\frac{N^2}{2X}(\eta^2 X^2-\omega^2)\frac{\partial V}{\partial X}\Bigr]\label{2-19}
\end{eqnarray}
\begin{eqnarray}
\ddot\omega=N^4\Bigl(\omega''+\frac{\omega'}{r}+2\frac{\omega'N'}{N}\Bigr)+2\frac{\dot \omega\dot N}{N}+\frac{2\omega}{(\eta^2 X^2+\omega^2)}\Bigl[N^4(\eta^2 X'^2+\omega'^2)\cr
-(\eta^2\dot X^2+\dot\omega^2)+N^2(V-\eta X\frac{\partial V}{\partial X}) -\frac{N^2}{2\omega}(\eta^2 X^2-\omega^2)\frac{\partial V}{\partial \omega}\Bigr].\label{2-20}
\end{eqnarray}
The superfluous equation for $\omega$ delivers a constraint equation,
\begin{eqnarray}
{N^\varphi}^2=\frac{1}{\eta^2 n^2 X^2}\Bigl[\frac{\eta^2 n^2 X^2}{r^2}+\eta X\frac{\partial V}{\partial X} +\omega\frac{\partial V}{\partial \omega} +V\Bigr],\label{2-21}
\end{eqnarray}
which is consistent with the requirement of the tracelessness of the energy momentum tensor\footnote{The tracelessness of the energy-momentum tensor can be seen as part of the field equations.}. This constraint is used in order to decouple the equations for $N^\varphi$.

We observe that the equations for the dilaton and scalar equations are identical, except the potential terms, as expected. They are of the type of a Klein-Gordon PDE.

Further, the equations for $X$ and $\omega$ are invariant under $t\rightarrow -t$ and $r\rightarrow -r$. So the antipodal pulse wave fulfills the same equation.
From the equations for   $N^\varphi $, it follows that the most realistic solution is  $N^\varphi =0$. The potential becomes
\begin{equation}
V=\alpha_1 X^{\alpha_2}\omega^{\frac{2}{3}-\eta \alpha_2}-\frac{3n^2\eta^2}{2r^2(3\eta -1)}(X^2+\alpha_3 X^{\frac{2}{3\eta}}),\label{2-22}
\end{equation}
with $\alpha_i$ constants.
Another solution is given by
\begin{equation}
N^\varphi =F_1(t)-\frac{3}{2}\int F_2(r)\log(\eta^2 X^2+\omega^2) dr,\label{2-23}
\end{equation}
with $F_1$ and $F_2$ arbitrary functions. It turns out that this solution is of less importance.
%========================================================================================
\subsubsection{case:  $A(t,r)=cst$.}
%=======================================================================================
A more interesting situation is obtained by taking a constant gauge potential $A_\varphi=-\frac{n}{e}$, i.e., $P(t,r)=0$ in Eq.(\ref{2-4}). The equations then becomes
\begin{eqnarray}
\ddot N=-N^4\Bigl(N''+3\frac{N'}{r}+\frac{N'^2}{N}\Bigr)+3\frac{\dot N^2}{N}
+\frac{2N}{(\eta^2 X^2+\omega^2)}\Bigl[y_0N^2V\cr
+3N^4(\eta^2X'^2+\omega'^2)-3(\eta^2\dot X^2+\dot\omega^2)-\frac{3}{4r}N^4(\eta^2 XX'+\omega\omega')\Bigr]\label{2-24}
\end{eqnarray}
\begin{eqnarray}
\ddot X=N^4\Bigl(X''+\frac{X'}{r}+2\frac{X'N'}{N}\Bigr)+2\frac{\dot X\dot N}{N}+\frac{N^2}{\eta}\frac{\partial V}{\partial X}\cr +\frac{2X}{(\eta^2 X^2+\omega^2)}\Bigl[N^4(\eta^2 X'^2+\omega'^2)-(\eta^2\dot X^2+\dot\omega^2)+\frac{y_0}{3}N^2 V\Bigr]\label{2-25}
\end{eqnarray}
\begin{eqnarray}
\ddot\omega=N^4\Bigl(\omega''+\frac{\omega'}{r}+2\frac{\omega'N'}{N}\Bigr)+2\frac{\dot \omega\dot N}{N}+y_0N^2\frac{\partial V}{\partial \omega}\cr +\frac{2\omega}{(\eta^2 X^2+\omega^2)}\Bigl[N^4(\eta^2 X'^2+\omega'^2)-(\eta^2\dot X^2+\dot\omega^2)+\frac{y_0}{3}N^2 V\Bigr].\label{2-26}
\end{eqnarray}
From the superfluous $\omega$ equation we obtain that the potential must be taken
\begin{equation}
V(X,\omega)=\beta_1 X^{\beta_2}\omega^{\frac{2}{3}-\frac{\eta\beta_2}{\vert y_0\vert}},\label{2-27}
\end{equation}
which is consistent with the tracelessness of the energy-momentum tensor\footnote{Remember that $\omega$ contains the gravitational constant $\kappa$ by the redefinition. }. Observe that the scale of the extra dimension enters the equations by $y_0$.
It is remarkable that we can obtain in the Lagrangian a quartic conformal invariant matter  coupling 
\begin{equation}
M\sim X^2\omega^2,\label{2-27b}
\end{equation}
(Eq.(\ref{2-27})) for suitable combination of the parameters, i.e.,
\begin{equation}
\vert y_0\vert =\frac{3}{2}\eta.\label{2-28}
\end{equation}

If we inspect the equations for the scalar and dilaton field (for $N=1$), we observe that the total potential is
\begin{equation}
V_X=\frac{\alpha \Bigl(\eta X^2(3\eta\beta+2y_0)+3\beta\omega^2\Bigr)}{3\eta(\eta^2 X^2+\omega^2)}X^{\beta -1}\omega^{\frac{(2y_0-3\beta\eta)}{3y_0}},\label{2-28b}
\end{equation}
\begin{equation}
V_\omega=\frac{\alpha \Bigl(\eta^2 X^2(2y_0-3\eta\beta)+\omega^2(4y_0-3\beta\eta)\Bigr)}{3(\eta^2 X^2+\omega^2)}X^{\beta}\omega^{-\frac{(y_0+3\beta\eta)}{3y_0}}.\label{2-28c}
\end{equation}

In fact, one can still integrate the functional integral over $\omega$ exactly. So non-conformal matter does not  effect the conformal invariance of the effective action after integrating over $\omega$ (see for example the discussion by ´t Hooft\cite{thooft2015b}).

The equations for  $N^\varphi$ are
\begin{eqnarray}
\dot {N^\varphi}'= -3{N^\varphi}'\frac{(\eta^2 X\dot X+\omega\dot\omega)}{(\eta^2X^2+\omega^2)}, \cr
{N^{\varphi}}''=-3{N^\varphi}'\Bigl[\frac{1}{r}+\frac{(\eta^2 X X'+\omega\omega')}{(\eta^2X^2+\omega^2)}  \Bigr].\label{2-29}
\end{eqnarray}
The general solution is
\begin{equation}
N^\varphi=f_1\int\frac{1}{r^3(\eta^2X^2+\omega^2)^{3/2}}dr,\label{2-30}
\end{equation}
with $f_1 $ a constant.
The scalar and dilaton equations differ only by the potential term. This difference is manifest in the numerical solution, presented in the next section.
Note that $n$ has disappeared from the PDE's. We have  absorbed  the  windingnumber (or vortex number) $n$ in the constant gauge field $\vert A_\varphi\vert=\frac{n}{e}$\footnote{The scalar-gauge system possesses  a quantized magnetic flux $\sim \frac{n}{e}$, which equals the first Chern number of $A$ on $\mathds{R}^2_+$. This has an important consequence in the expansion of the scalar field in cylindrical harmonics. See section 4.}
\begin{figure}[h]
\centerline{
\includegraphics[width=0.35\textwidth]{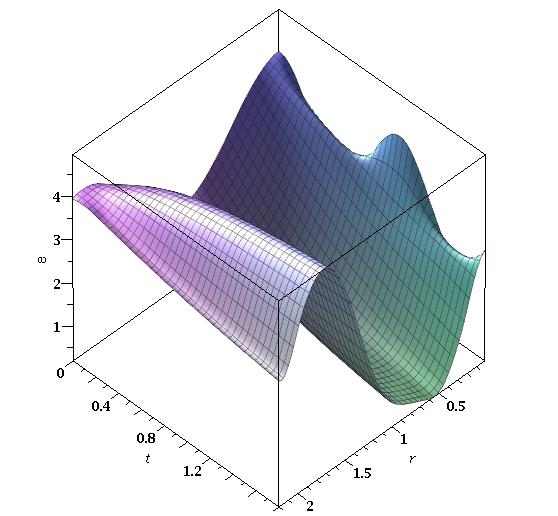}
\includegraphics[width=0.35\textwidth]{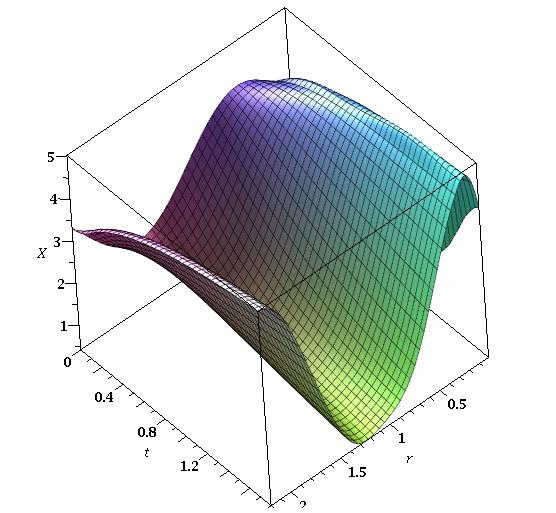}
\includegraphics[width=0.4\textwidth]{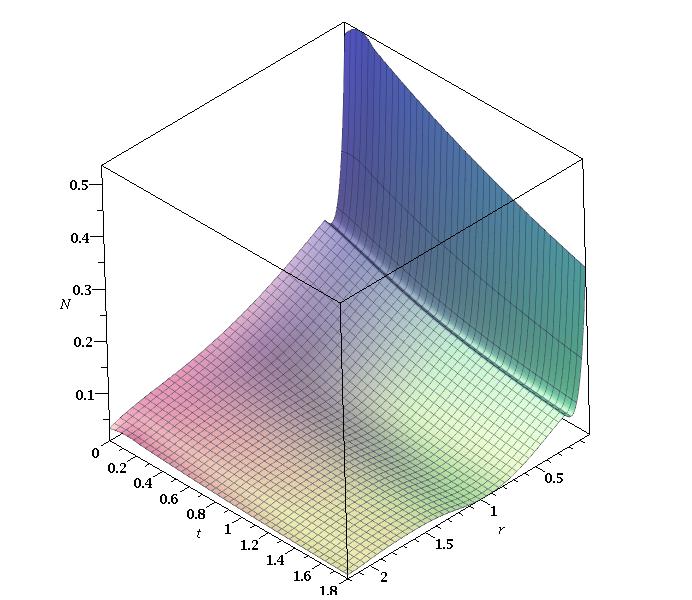}}
\caption{stable solution  of $X, \omega$ and $N$ for the case of a constant gauge field. The potential is $V=\beta_1 X^{\beta_2}\omega^{\frac{2}{3}-\frac{\eta\beta_2}{y_0}}$. We took for the initial values of the  scalar and dilaton field a cosine and sine function respectively. For  $N$ we took the vacuum solutions of the Appendix A. Further, we applied Neumann boundary conditions. Note that  $N$ develops a singularity and approaches to a constant value for increasing time. The solution depends critically on the parameters of the potential, for example, the scale $y_0$ of the extra dimension. }\label{fig1}
\end{figure}
%=====================================================================================
%======================================================================================
\subsubsection{The numerical solution}
%======================================================================================
It is clear that we can have breather modes of our scalar-dilaton system.
In figure 1 and 2 we plotted a typical solution in the case of a constant gauge field.
%=========================================================================================
\begin{figure}[h]
\centering
\includegraphics[width=0.34\textwidth]{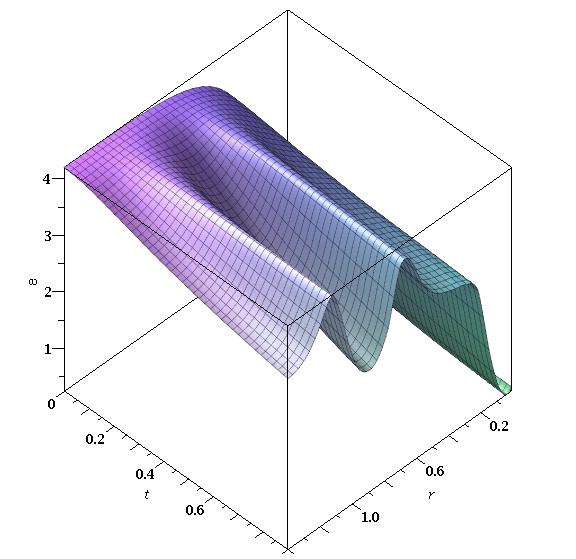}
\includegraphics[width=0.34\textwidth]{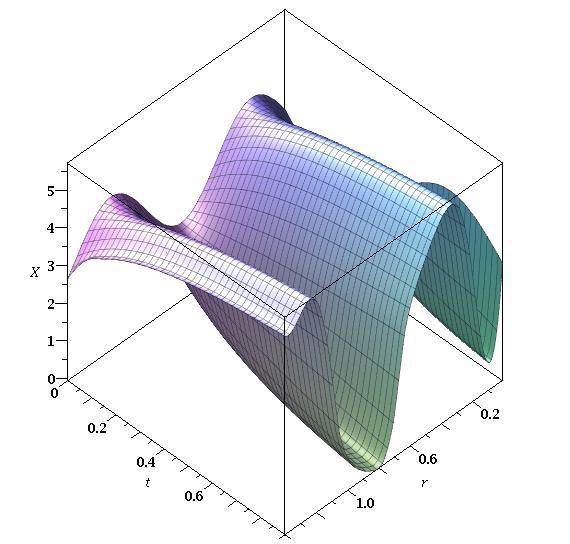}
\includegraphics[width=0.34\textwidth]{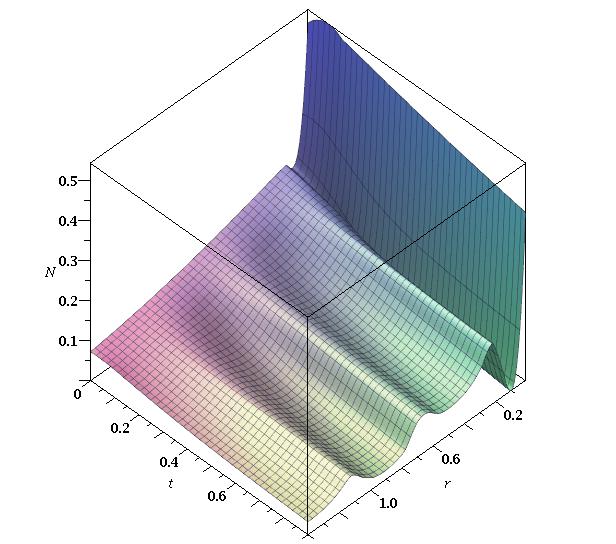}
\includegraphics[width=0.37\textwidth]{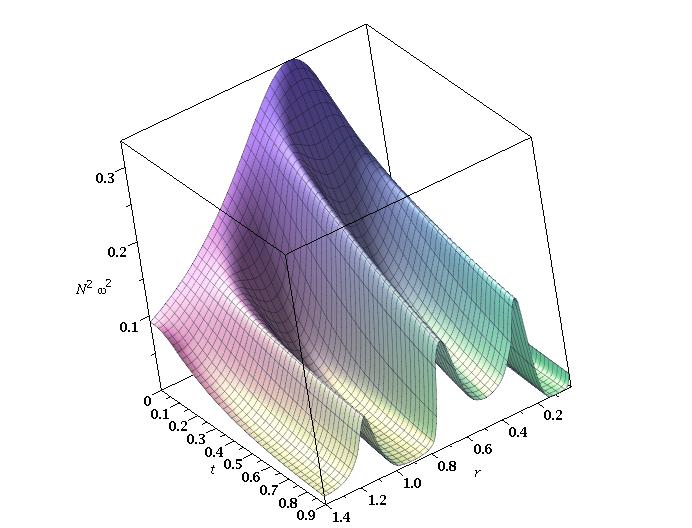}
\caption{As figure 1, with now as initial value for $\omega$ a $\tanh$ function. We observe that the polynomial behavior in the metric component $N$  is induced by the dilaton. We also plotted the total metric component $\omega^4N^2$}\label{fig2}
\end{figure}

%===========================================================================
\section{On the double cover of the Klein surface, Riemann surfaces and antipodicity }\label{sec3}
%===========================================================================
\subsection{Motivation for the Klein surface}
%==========================================================================
For the $S^2$ we could apply the stereographic projection $S^2\rightarrow \mathds{R}P^2\rightarrow \mathds{C}P^1$. In order to make the map one-to-one, one applies the $\mathds{Z}_2$ symmetry identification: the two antipodal planes are identified, $z\rightarrow -\frac{1}{\bar z}$ (see figure 3).
In our situation, we uplift the projection to $S^3\subset \mathds{R}^4$  and use cylindrical coordinates $(r,\varphi)$.
\begin{figure}[h]
\centering
\includegraphics[width=0.45\textwidth]{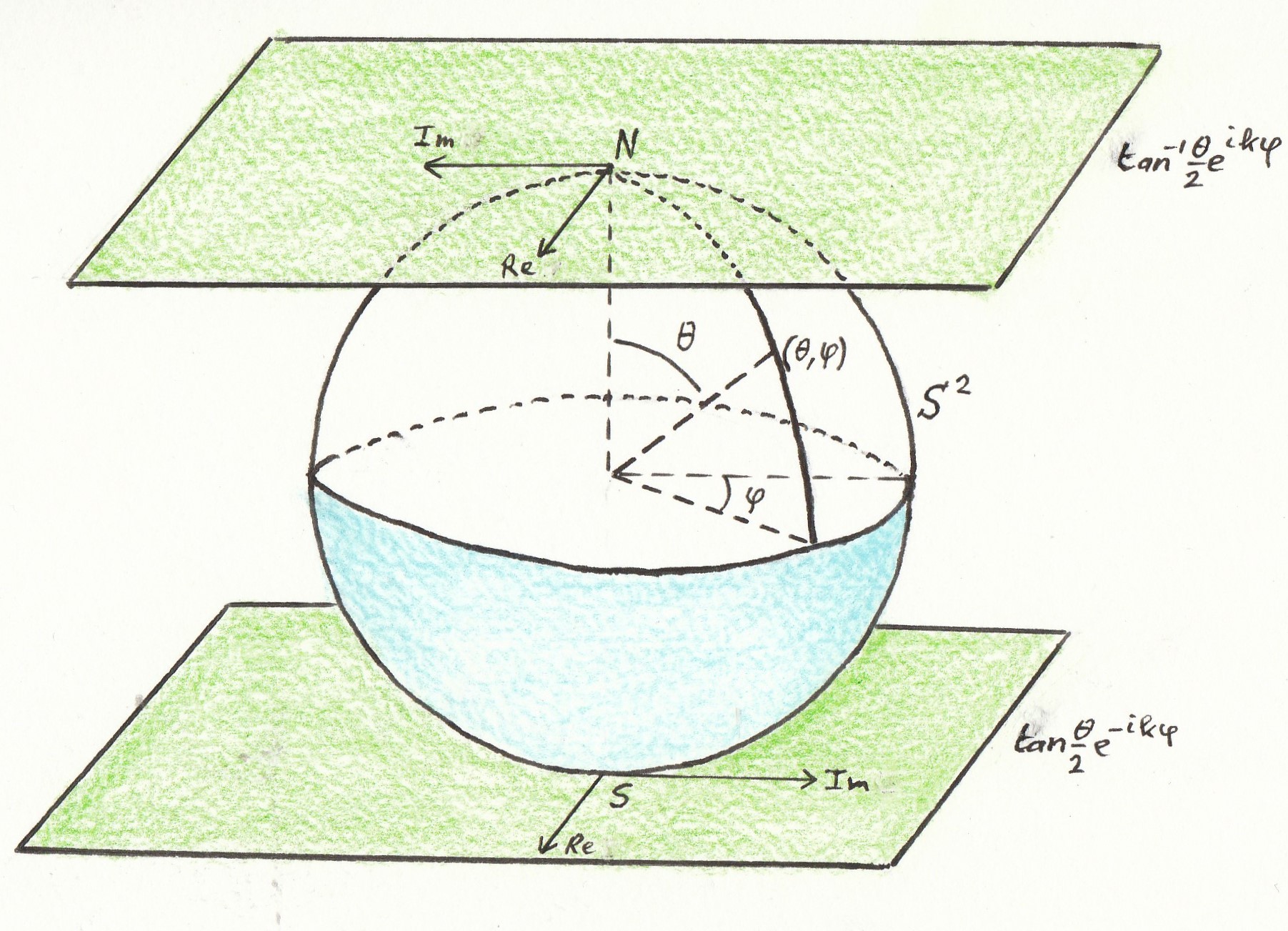}
\includegraphics[width=0.47\textwidth]{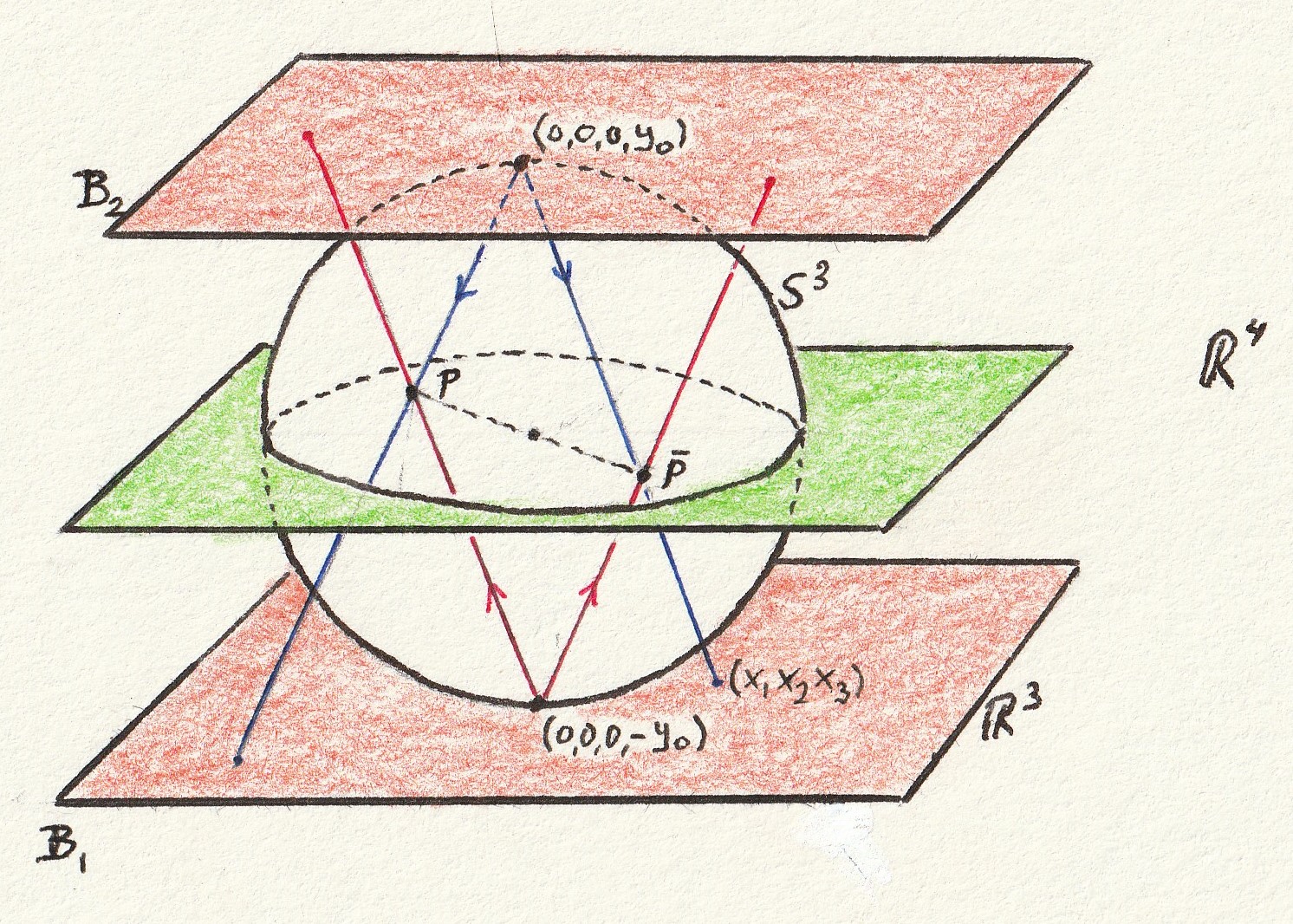}
\caption{Left: the stereographic projected $S^2\rightarrow \mathds{R}P^2\rightarrow \mathds{C}P^1$. In order to make the map one-to-one, one applies the $\mathds{Z}_2$ symmetry identification: the two antipodal planes are identified, $z\rightarrow -\frac{1}{\bar z}$. 
Right: stereographic projection of $S^3\subset \mathds{R}^4\simeq \mathds{C}\times\mathds{C}$ with also antipodal identification, representing the projection of an embedded Klein bottle.}\label{fig2}
\end{figure}
\begin{figure}[h]
\centering
\includegraphics[width=0.4\textwidth]{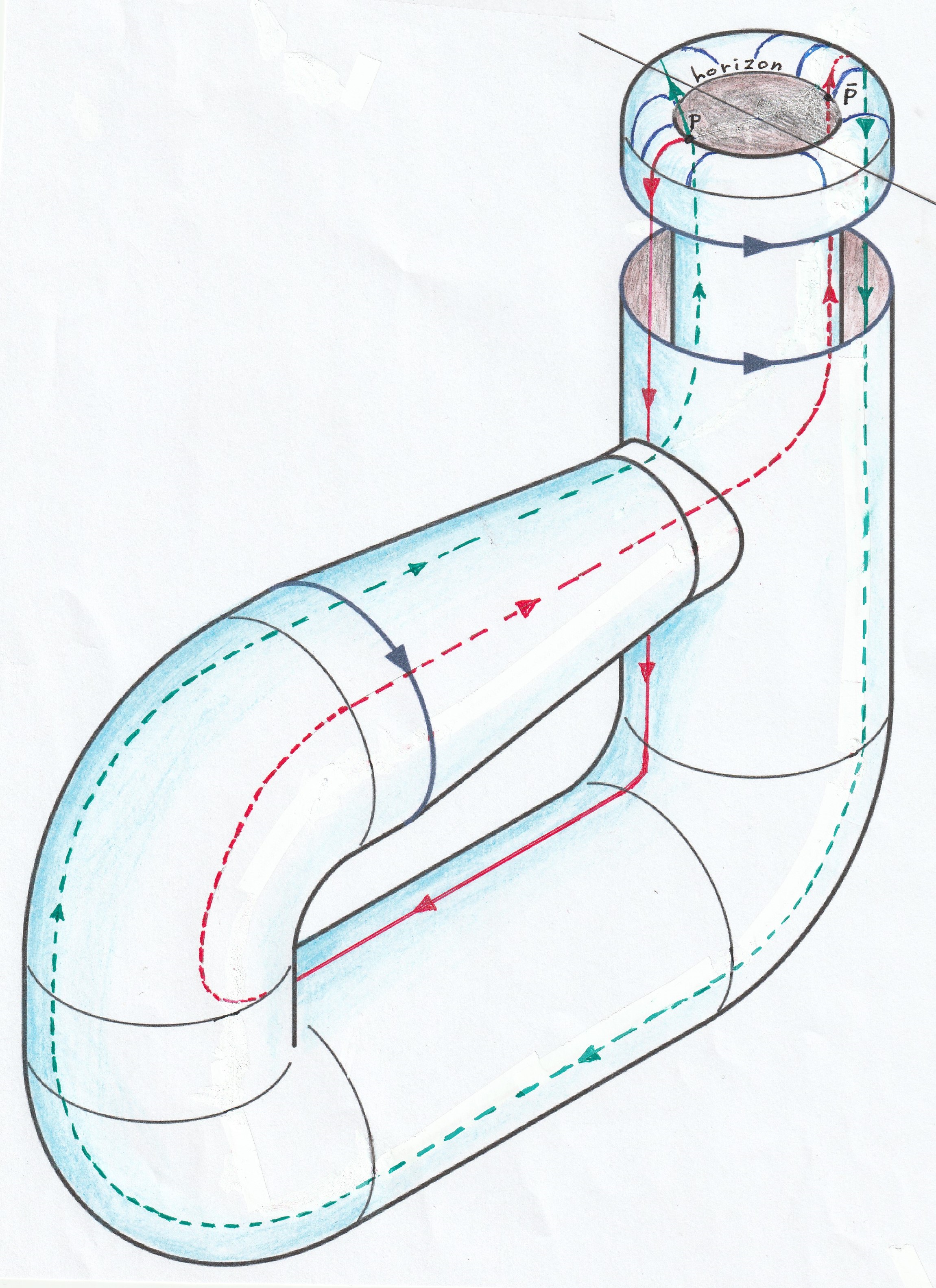}
\includegraphics[width=0.4\textwidth]{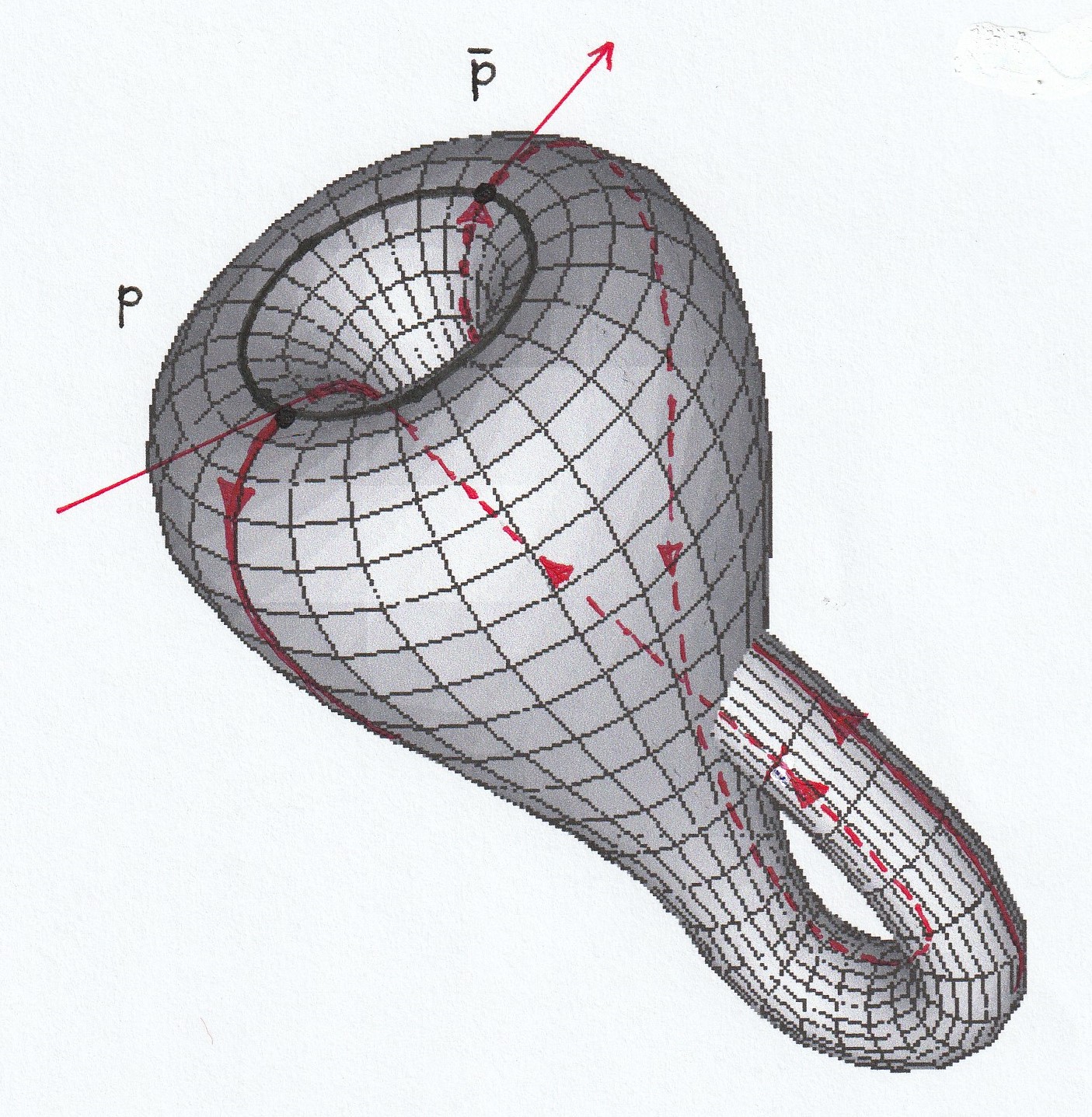}
\caption{Artist impression of the trajectory of a scalar (Hawking-) particle  on the Klein surface. $P$ and $\bar P$ are the antipodes. Again, we have a $\mathds{Z}_2$ symmetry identification (see figure 3), now for the both sides of the brane: $g_{\mu\nu}({\bf x},\mathpzc{y})=g_{\mu\nu}({\bf x},\mathpzc{-y})$. Note that we apply a slightly different antipodal mapping. We are dealing here with cylindrical symmetric Kerr-like spacetime, where we expand the scalar field in cylindrical harmonics $\Phi_{n,m}=\eta X(r,t)e^{in\varphi}Y_m(\Omega)$, in stead of spherical ones. Further, we replace the "cut-and-paste" method by the antipodicity in the bulk space. We can bypass the notion of "quantum clones" proposed by ´t Hooft recently\cite{thooft2022}.} \label{fig2}
\end{figure}
By means of the Hopf fibration, one then makes the connection with real spacetime. For details, we refer to Slagter\cite{slagter2022c} and Urbantke\cite{ur2003}.
Another important characteristic of the non-orientable Klein surface is the fact that meromorphic functions remain constant\cite{alling1971}. Our solution for $N$ is meromorphic\footnote{Remember that our solution for $N$ of section 4.2 (or see Appendix A) could be written as\cite{slagter2022c} as a meromorphic polynomial $\frac{P({\bf z})}{Q({\bf z})}$, with $Deg(P)=5$. }. Further, the Klein surface is homeomorphic to the connected sum of two projective planes.

In the past, intensive research was done on the fundamental group structure of compact surfaces in $m$ dimensions. It is clear that, for example, the torus possesses the structure of two infinite cyclic groups. For the projective plane, the cyclic group of order 2. For the Klein bottle, the fundamental group can be presented by two generators $a,b$ and $baba^{-1}$\cite{massey1971}.

We constructed our model in the 5D RS warped spacetime. So the question is how we can imagine the evolution of the Hawking particles created near the 
horizon, say point $P(U,V,z,\varphi, \mathpzc{y})$ in figure 4. The in-going particle can travel on the Klein surface, in order to leave in the antipode $\bar P(-U,-V,-z,\varphi+\pi ,-\mathpzc{y})$ the horizon ($J:P\rightarrow \bar P$ is an inversion in 5D).

The real physics still takes place on the brane. Effectively, we replaced the "cut-and-past" method in $\mathds{R}^3$ by the imaginary transport in $\mathds{R}^4$ (via the Hopf fibration of the 3-sphere in $\mathds{R}^4$.

We know that complete embedded surfaces in $\mathds{R}^3$ must be orientable, otherwise we have self-intersection, so the genus changes.
Non-orientable surfaces immersed in $\mathds{R}^3$ does not have global Gauss maps, so no well defined mean curvature.
So the application of the Klein surface embedded in $\mathds{R}^4$ is quite interesting in our context.
First, we complexify our hyper-surface\cite{slagter2022c}.
Let us  write our coordinates $( r, z, \mathpzc{y},\varphi^*)$ as\footnote{with $x=r\sin\varphi, y=r\cos\varphi$,}
\begin{equation}
{\cal V}=z+i\mathpzc{ y}=Re^{i\varphi_1},\qquad {\cal W}=x+iy=re^{i\varphi_2},\label{3-1}
\end{equation}
where the antipodal map is now ${\cal V}\rightarrow -{\cal V}\equiv-R^2/\bar{\cal  V} , {\cal W}\rightarrow -{\cal W}\equiv-r^2/\bar{\cal  W}$.\footnote{ because $e^{i(\varphi+\pi)}=-e^{i\varphi}$. }
Further, ${\cal V}\bar{\cal V}=z^2+\mathpzc{y}^2=R^2,{\cal W}\bar{\cal W}=x^2+y^2=r^2 $.
And after inversion
\begin{eqnarray}
z=\frac{1}{2}({\cal V}+\bar{\cal V}), \qquad \mathpzc{y}=\frac{1}{2i}({\cal V}-\bar{\cal V}),\hspace{0.5cm}\cr
x=\frac{1}{2}({\cal W}+\bar{\cal W}),\qquad y=\frac{1}{2i}(({\cal W}-\bar{\cal W}),\cr
\varphi_1 =i\log\sqrt{\frac{\bar{\cal W}}{{\cal W}}}\quad \varphi_2= i\log\sqrt{\frac{\bar{\cal V}}{{\cal V}}}.\label{3-2}
\end{eqnarray}
We can write   
\begin{equation}
dz^2+d\mathpzc{y}^2+dx^2+dy^2=d{\cal V}d\bar{\cal V}+d{\cal W}d\bar{\cal W}.\label{3-3} 
\end{equation}
We now have $\vert{\cal V}\vert^2 +\vert{\cal W}\vert^2=x^2+y^2+z^2+\mathpzc{y}^2=r^2+R^2$.
We identified $\mathds{C}^1\times\mathds{C}^1$ with $\mathds{R}^4$ and so contains $S^3$, given by $\vert {\cal V}\vert ^2 +\vert{\cal W}\vert^2=const.$ 
Every line through the origin, represented by $({\cal V}, {\cal W})$ intersects the sphere $S^3$, for example $(\lambda{\cal V}, {\lambda\cal W})$ with $\lambda=\frac{1}{\sqrt{\vert {\cal V}\vert ^2 +\vert{\cal W}\vert^2}}$. Thus the homogeneous coordinates can be restricted to $\vert {\cal V}\vert ^2 +\vert{\cal W}\vert^2=1.$ 
The point  $({\cal V}, {\cal W})\in S^3\subset\mathds{C}^1\times\mathds{C}^1$ with $\vert {\cal V}\vert ^2 +\vert{\cal W}\vert^2=1$, becomes by the complexification,  a point of $S^2$, so with the single complex coordinate ${\cal Z}=\frac{{\cal V}}{{\cal W}}$. We have now a map $H:S^3\rightarrow S^2$, which is continuous. 
One calls this a  Hopf map.
For each point of $S^2$, the coordinate $({\cal V}, {\cal W})$ is non unique, because it can be replaced by $(\lambda{\cal V}, \lambda{\cal W})$, such that $\vert\lambda\vert^2=1, \lambda\in S^1$.
SWe will now write $\mathds{C}_1$ for $S^2/\{\infty\}$ and $\mathds{C}_2$ for $S^2/\{0\}$ and admitting coordinates ${\cal Z}$ and ${\cal Z}'=\frac{1}{{\cal Z}}$ respectively\footnote{Let $G$ be the group of self-homeomorphisms of the product space $S^2\times S^2$, generated by interchanging the two coordinates of any point and by the antipodal map on either factor. 
$G$ is then isomorphic to the  dihedral group. It contains 3 subgroups, for example $K=\{I,(x,y)\rightarrow (-x,y), (x,y)\rightarrow (x,-y), (x,y)\rightarrow (-x,-y)\}$. It acts freely on $S^2\times S^2$. Then $(S^2\times S^2)/K=\mathds{R}P^2\times\mathds{RP}^2$.
The most interesting feature is the fact that the 2-fold symmetric product of $\mathds{R}P^2$,  $SP^2 (\mathds{R}P^2)=\mathds{R}P^4$.}.

If we write 
\begin{equation}
{\bf H} = 
\begin{pmatrix}
z+i\mathpzc{y}& x+iy \\
-x+iy & z-i\mathpzc{y} 
\end{pmatrix}=
\begin{pmatrix}
\cal{V}& \cal{W} \\
-\cal{\bar W} &\cal{\bar V} 
\end{pmatrix}\label{3-4}
\end{equation}
with $det({\bf H})=\vert{\cal V}\vert^2+\vert{\cal W}\vert^2$ and $\mathds{R}^4\cong {\bf H}$,
we can describe the mapping also with normalized quaternions $\mathds{A}=[a,{\bf A}]=[\cos\alpha,\sin\alpha {\bf n}]$ as binary rotations\cite{altman1986}. This is clarifying, when we
consider the covariance covering  groups, such as the $SO(3)$ and  $SU(2)$.
A rotation is actually represented by two antipodal points on the sphere (a kind of double cover) as a point on the hypersphere in 4D. Each rotation represented  as a point  on the hypersphere is matched by its antipode on that hypersphere. The quaternion represents a point in 4D. Constrained to unit magnitude, yields then a 3D space, i.e., the surface of a hypersphere.
The group of unit quaternions $S^3\subset{\bf H}$ is isomorphic with $SU(2)$.
We identify $S^3$ and $SU(2)$ with the isomorphism. The relation with the M\"obius group $G$ is then easily made (see Toth\cite{toth2002} or Slagter\cite{slagter2022c}) and  so with the alternating group  ${\cal A}_5$, isomorphic with the binary symmetry group of the icosahedron (see for example\cite{toth2002}). 
Because the icosahedron can be circumscribed by 5 tetrahedra, they form an orbit of order 5 symmetry rotations of the icosahedron. These symmetries are subgroups of the icosahedron symmetry group. 
The connection with our quintic solution was made in a former study\cite{slagter2022c}, by the observation that the vertices of the icosahedron, stereographically projected to $\mathds{C}$, are
\begin{equation}
0, \infty, \gamma^j(\gamma+\gamma^4), \gamma^j(\gamma^2+\gamma^3), \quad j=0,..4,\quad \gamma=e^{\frac{2\pi i}{5}}.\label{3-5}
\end{equation} 
Shortly stated, for the icosahedral M\"obius group, by suitable orientation of the axes, we can write the linear fractional transformations as $\zeta\rightarrow \gamma^m\zeta (m=0..4)$.
The binary icosahedral group $G^*$ associated with the M\"obius group $G\subset{\cal M}_0(\mathds{C})$\footnote{Remember, $SO(3)\cong{\cal M}_0(\mathds{C})=SU(2)/ \{\pm I\}$.}.

Let us now try to make the connection with state vectors  in $\mathds{C}^2$.
On a complex Hilbert space ${\cal H}$, taken as the space $\mathds{C}^2$ of pairs $\ket{{\cal U}}=({\cal V},{\cal W})$ and equipped with scalar product, one can define a state vector by the set of multiples $\lambda\ket{{\cal U}} $, with $\vert\lambda\vert=1$ and normalization condition  $\braket{{\cal U}}{{\cal U}}=1$. Further, we have the matrix 
\begin{eqnarray}
\rho:=\ket{{\cal U}}\bra{{\cal U}},\quad 
\rho_{ij}=\begin{pmatrix}
\cal{V}\cal{\bar V}& \cal{V}\cal{\bar W} \\
\cal{ W}\cal{\bar V} &\cal{W}\cal{\bar W} 
\end{pmatrix},\label{3-6}
\end{eqnarray}
with $\rho^\dag=\rho$ and ${\bf Tr}(\rho)=1$.
We define now the vector $\vec{{\bf V}}=(V_1,V_2,V_3)\in \mathds{R}^3$, with $\rho_{11}=\frac{1}{2}(1-V_3), \rho_{22}=\frac{1}{2}(1-V_3)$ and $\rho_{21}=\bar\rho_{12}=\frac{1}{2}(V_1+iV_2)$.
We can write $\rho=\frac{1}{2}(\mathds{I}+\vec{{\bf V}}.{\bf\sigma})$, with $\sigma$ the Pauli matrices. Furter, $1-{\bf V}^2\geq 0$.
We define the vector $\vec{{\bf X}}=(x,y,z,\mathpzc{y})$ and take $\lambda=e^{i\alpha}$ a phase factor. We then write ${\cal W}_i\rightarrow e^{i\alpha}{\cal W}_i, {\cal V}_i\rightarrow e^{i\alpha}{\cal V}_i$. So $\vec{{\bf X}}\rightarrow \cos(\alpha)\vec{{\bf X}}+\sin(\alpha)J\vec{{\bf X}}$, with $J\vec{{\bf X}}=(-X_2, X_1, -X_4, X_3)=(-y,x,-\mathpzc{y},z)$ a normalized combination of two orthogonal unit vectors of $\mathds{R}^4$ for all $\alpha$. In fact, we obtain a Hopf fibration of $S^3$. This fibering is the stereographic projection on $X_4=\mathpzc{y}=0$ of $S^3$, an analogy of the stereographic projection of one dimension lower. See figure 3. Remember, we identified the antipodal points, so we obtain a Klein bottle $\mathds{K}=\{({\cal V},{\cal W});\vert{\cal V}\vert^2-\vert{\cal W}\vert^2=\cos(\theta) \}\in S^3\subset\mathds{C}^2 $ in $\mathds{R}^3$ (see Eq.(\ref{3-7})). However, it is embedded in our $\mathds{R}^4$.
If we project from $(0,0,0,\pm\mathpzc{y}_0)$, the running point $\vec{{\bf X}}$ on $S^3$ is now related to its stereographic image $\vec{{\bf x}}=(x_1,x_2,x_3)$ by
$X_4=\pm\frac{{\bf x}^2-1}{{\bf x}^2+1}$, $X_i=\frac{2x_i}{{\bf x}^2+1}$.
For normalization of $\ket{{\cal Z}}$, we first define $\varphi =\varphi_2-\varphi_1$ and write
\begin{eqnarray}
{\cal W}=\cos(\frac{\theta}{2})e^{i\varphi_1}=\cos(\frac{\theta}{2})e^{-i\varphi/2} e^{i(\varphi_1+\varphi_2)/2} \cr
{\cal V}=\sin(\frac{\theta}{2})e^{i\varphi_2}=\sin(\frac{\theta}{2})e^{i\varphi/2} e^{i(\varphi_1+\varphi_2)/2}.\label{3-7}
\end{eqnarray}
Varying $\alpha$  will only change $(\varphi_1+\varphi_2)/2$ and not $\varphi$ and $\theta$. Further we have $V_3=\cos(\theta)$ and $V_1+iV_2=\sin(\theta)e^{i\varphi}$, ${\cal Z}=\frac{{\cal V}}{{\cal W}}=\tan(\theta/2)e^{i\varphi}$, with the antipodal identification $\varphi\rightarrow \varphi+\pi$.

One can compare this visualization by the orientable counterpart situation of the torus. For details, see Toth\cite{toth2002} (section 1.4) or Urbantke\cite{ur1990}

So if we consider a Hawking particle falling in, it travels for  a while in $\mathds{R}^4$ from $\mathpzc{y}_0\rightarrow -\mathpzc{y}_0$ arriving at the antipodal point in $\mathds{R}^3$.

%===========================================================================
\subsection{The elapse time in the bulk}
%=============================================================================
One could wonder if one can estimate the elapse propertime of a particle when it moves in the bulk space and back to the brane\footnote{Another interesting  method could be delivered by the equation of the directrix of the Klein bottle. From the integral curve from the PDE's, $\frac{\partial r}{\partial t_e}$, one could find, in principle, the elapse time\cite{massey1971}.}.
Our horizon $r_H$ is determined by  $a_i$ (see Eq.(\ref{A-4})).
We assume that $y_0\sim 10^{-5}m$\cite{maartens2010} and let us take $r_H\sim 10^{7}$m. See, for example, Maldacena\cite{maldacena2021}\footnote{There is a difference between Maldacena's method and our model: we don't need a matter field in the bulk.}. 
The proper traversal time, $t_e$, of the Klein surface will be of the order of $r_H$. Further, the trajectory on the Klein surface in the bulk, $l_b$, will be of the order $\frac{t}{t_e}$, with $t$ the asymptotic time and $t_e$ the proper time. 
If we adapt the approximation of the wormhole model on the RS spacetime of Maldacena, we could obtain an elapse time of the order $10^{-2}$ s, while $t$ would be must larger.

It would be of interest if one could measure this elapse time in the Hawking radiation. One could obtain an estimate of $y_0$.
%===========================================================================
\subsection{The scalar equation on the single-sided black hole}
%==================================================================================
It is obvious that strong gravitation interactions will take place near the horizon. The scattering process by in- and out-going wave functions can be handled  by different approximations.
One can make a distinction between the near horizon behavior and the further away approximation by means of the Regge-Wheeler potential\cite{betzios2021}. An effective gravitational potential for the propagating modes on a Schwarzschild background then enters the Klein-Gordon equation.

We will consider here the antipodal boundary condition of  the single-sided black hole. This idea was already studied by Schr\"odinger\cite{schrod1957}. It was used by Gibbons\cite{gib1986} in connection with CPT invariance and quantum mechanics on the identified space. The method was also used by Sanchez, et al.\cite{san1987} in connection with the quantum Fock space.
We applied the mapping on the RS spacetime\cite{slagter2022c}. We already observed that the scale of the extra dimension $y_0$ entered the effective 4D field equations. Further, the dynamical evolution of the 4D Einstein equations were modified by the projection of the 5D Weyl tensor on the brane and carried the information of the gravitation field outside the brane (the so-called "Kaluza-Klein"modes).

We will use the expansion of the scalar field in cylindrical harmonics\footnote{We work in the axially symmetric spacetime. Moreover, our scalar gauge field is also rotational symmetric, a fundamental property of the vortex, which contains the winding number n, which is a topological parameter.} 
\begin{equation}
\Phi_{n,m}=\eta X(r,t)e^{in\varphi}{\bf Y}_m,\label{3-8}
\end{equation} 
where $X(r,t)$ fulfills the PDE's of section 2. Further, we have $e^{in(\varphi +\pi)}=(-1)^n e^{in\varphi}$.
For ${\bf Y}_m$ we can choose 
\begin{equation}
{\bf Y}_m=\sum_{m}\Bigl(A_m\cos(m\varphi)+B_m\sin(m\varphi)\Bigr).\label{3-9}
\end{equation}
We can in this way always obtain a factor $(-1)^{(n+m)}$ in front of the right-hand side in Eq.(\ref{3-8}). Further, the field equations are invariant under $t\rightarrow -t$ (see 4.2 and the Appendix A).
Note that for antipodicity, we need  the map: $e^{i\varphi}\rightarrow e^{i(\varphi +\pi)}$.

The reader must remind himself that in our conformal model, the scale is determined by the dilaton $\omega$. In the vacuum situation (see Appendix A) an exact time-dependent solution was found. 
Now we conjecture that in the non-vacuum situation, more information can be gathered by our model about the scattering process when  approaching smaller scales. The distinction one usually makes between the near horizon region and the Regge-Wheeler region, can be replaced by the dynamical interaction of the scalar waves and the dilaton field. The distinction was necessary because gravity comes into play. 
Further, the small scale behavior translates itself in the dilaton and the conformally flat $\tilde g_{\mu\nu}$. 
Remember now that the dilaton field (part of the gravitation metric) describes the complementarity between the local and outside observer and 
can be used (by $\Omega$, see Eq.(\ref{2-2})) to define the vacuum state the local observer experiences (see section 2.1).
It is conjectured (also from the numerical solutions) that the spacetime becomes singular-free when $t\rightarrow\infty$,  except for $r=0$, which is no problem because there is no inside of the Klein surface.
%========================================================================
\subsection{Comparison with AdS}
%=======================================================================
The elliptic interpretation was historically first studied in cosmology, i.e., in the deSitter spacetime.
Let us consider the inversion in $\mathds{R}^5, J: {\bf X}^a\rightarrow -{\bf X}^a, (a=1...5)$, and the hypersurface $\frac{1}{H^2}=-t^2+x^2+y^2+z^2+\mathpzc{y}^2$, which is  the deSitter spacetime in a 5D spacetime\footnote{Note that we consider in our model the topology of the Klein surface.}
\begin{equation}
ds^2=-dt^2+dx^2+dy^2+dz^2+d\mathpzc{y}^2,\label{3-10}
\end{equation}
with $H^{-1}$ the radius of the hypersphere and $t$ the proper time along a set of geodesics. The symmetries of this pseudo sphere  is the 10 parameter Poincar\'e group in Minkowski spacetime. One writes the pseude sphere ($\eta=-H^{-1}exp(-Ht)$)
\begin{equation}
ds_{dS}^2=-dt^2+e^{2Ht}(dx^2+dy^2+dz^2)=\frac{1}{H^2\eta^2}(-d\eta^2+dx^2+dy^2+dz^2).\label{3-11}
\end{equation}
The group invariance of the deSitter spacetime can now be used to apply a high-frequency behavior of the scalar field in order to define the prefered vacuum state as measured by the observers on the geodesics.
From the scalar equation (like Eq.(\ref{2-13})), one easily find
\begin{equation}
\partial_t^2\phi+3H\partial_t\phi-e^{-2Ht}\sum_{i}\partial_i^2\phi+M^2\phi=0,\label{3-12}
\end{equation}
with $M^2=m^2+12\xi H^2$. We expand the field operator $\phi$ now as
\begin{equation}
\phi=\sum_{k}\Bigl\{A_{\vec{k}}f_{\vec{k}}(x)+A^\dagger_{\vec{k}}f^*_{\vec{k}}(x)  \Bigr\},\quad f_{\vec{k}}=\frac{1}{\sqrt{2Ve^{3Ht}}}2^{i\vec{k}.\vec{x}}h_k,\label{3-13}
\end{equation}
with $k^i=2\pi n^i/L, n^i$ an integer, $k=\vert\vec{k}\vert$ and $L$ the dimension of the 3-space ($V\sim L^3$). The term $V e^{3Ht}$ represents here the physical volume of the cube on which one applies the periodic boundary conditions. Further $h_k$ satisfies  a Bessel equation
\begin{equation}
v^2\frac{d^2}{dv^2}h_k+v\frac{d}{dv}{h_k}+(v^2-\nu^2)h_k=0,\quad \nu=\sqrt{\frac{9}{4}-\frac{M^2}{H^2}},\quad v=\frac{ke^{-Ht}}{H}.\label{3-14}
\end{equation}
One can also use $u\equiv -v$ as independent variable. $v$ is the ratio of the momentum of a particle to the expansion $H$. In this context, an observer on a geodesic in the deSitter spacetime will be surrounded by isotropic thermal radiation ( with wavelength $\lambda$) coming from de geodesic's Hubble horizon with temperature $T_H=H/(2\pi k_B)$. So $v$ represents also the ratio of the proper radius $H^{-1}$  to $\lambda/(2\pi)$. For small wavelength, compared with the radius of curvature and slowly changing of $a(t)=e^{Ht}$\footnote{we write in general, $ds^2=a(t)^2(-dt^2+dx^2+dy^2+dz^2)$.}, we assume that the positive frequency solutions $f_{\vec{k}}$ possesses an adiabatic form, i.e., the WKB-solution. We can write
\begin{equation}
f_{\vec{k}}({\bf x})=\frac{1}{\sqrt{2Va^3\omega_k}}e^{i(\vec{k}.\vec{x}-\int\omega_kdt)},\quad \omega=k/a(t).
\end{equation}\label{3-15}
Further, $h_k$ has the asymptotic form 
\begin{equation}
h_k\sim\frac{1}{\sqrt{\omega}}e^{-i\int\omega_k dt}.\label{3-16}
\end{equation}
For large $k, h_k\sim \frac{1}{\sqrt{Hv}}e^{iv}$.
The general solution of the Bessel equation, Eq.(\ref{3-14}), is
\begin{equation}
h_k=\sqrt{\frac{\pi}{2H}}\Bigl\{E(k)H_\nu^{(2)}(v)+F(k)H_\nu^{(1)}(v)\Bigr\},\label{3-17}
\end{equation}
with $H_\nu$ Hankel functions. 
When $k\rightarrow \infty$, we have $E(k)\sim 0, F(k)\sim 1$. From the deSitter symmetries, one finally obtains the creation and annihilation operators
\begin{equation}
f_{\vec{k}}=\sqrt{\frac{\pi}{4HV}}e^{-3/2Ht}H_\nu^{(1)}(k\frac{k}{H}e^{-HT})e^{i\vec{k}.\vec{x}},\label{3-18}
\end{equation}
and hence the vacuum state\footnote{Also called the Bunch-Davies vacuum, valid also for the conformally coupled massless scalar. However, in our case, we could add the mass term of Eq(\ref{2-27b})}. For $\xi=1/6$ and $ m=0$, we obtain
$H_{\frac{1}{2}}^{(1)}=-i\sqrt{\frac{2}{\pi v}}e^{iv}$. For $u=-v, H_\nu^{(1)}(v)$ is replaced by $H_\nu^{(2)}(u)$.
Note that in calculating the expectation values such as $<\phi(\vec{x},t)^2>$,  there appears infrared or ultra violet singularities in this two point function, at least for $\xi=0$.

In the antipodal situation of the deSitter spacetime, the expectation value becomes (symmetric and antisymmetric)\cite{fol1987}
\begin{equation}
<\phi_{JA/S}(\vec{x},t)^2> = <\phi(\vec{x},t)^2>\pm<\bar \phi(\vec{x},t)^2>,\label{3-19}
\end{equation}
with 
\begin{equation}
<\bar\phi(\vec{x},t)^2>=\frac{1}{16\pi\cos(\pi\nu)}\{m^2+12(\xi-1/6)H^2\}.\label{3-20}
\end{equation}
What remains is the problem with the Fock vacuum. This is essential, because one can always add an antipodal source (see discussion by Gibbons\cite{gib1986}).

We now return to our situation. In the next sections we investigate some topological aspects.
%===========================================================================
\section{Topological aspects: relation with the instanton and minimal embedding in $\mathds{K}$}\label{sec4}
%===========================================================================
\subsection{Some history}
%=======================================================================
It is well known that on Minkowski spacetime $\mathds{R}^{d+1}$, the action of the static  Yang-Mills-Higgs  configuration $({\bf A},\Phi)$ 
equals the Euclidean action on $\mathds{R}^d$. 
The field equations of the Euclidean action, for example, in $\mathds{R}^4$ in  this model, allow soliton solutions, which are so time-independent finite energy solutions to the variational equations for the action density on the $\mathds{R}^5$ Minkowski spacetime. 
For the $d=4$ pure Yang-Mills case, they are called instantons. It turns out that their "curvatures" are self-dual\footnote{If one allows a Higgs field, then for $d=2$ one calls the configuration a vortex and for $d=3$ a monopole.}.

In the "transcendental" book of Jaffe and Taubes\cite{jaf1981} one finds a nice introduction of these issues.
We can now make the connection with our model.
%========================================================================
\subsection{Connection with the warped spacetime solution}
%=======================================================================
Let us write our (4+1) dimensional spacetime on the 5 dimensional  Riemannian manifold:
\begin{equation}
ds^2=\omega(\tau,r,y)^{4/3}\Bigl[N(\tau,r)^2d\tau^2+\frac{1}{N(\tau,r)^2}dr^2+dz^2+r^2(d\varphi+N^\varphi(\tau,r)d\tau)^2+d\mathpzc{y}^2\Bigr].\label{4-1}
\end{equation}
One easily obtains the field equations in $d$ ($d=4,5$; compare with Eq.(\ref{A-1}) and Eq.(\ref{A-2}) in the pseudo-Riemannian spacetime)
\begin{equation}
\ddot\omega =N^4\omega''-\frac{d}{\omega(d-2)}(N^4\omega'^2-\dot\omega^2),\label{4-2}
\end{equation}
\begin{eqnarray}
\ddot N=\frac{3\dot N^2}{N}+N^4(N''+\frac{3N'}{r}+\frac{N'^2}{N})\cr
+\frac{(d-1)}{ (d-3)\omega}\Bigl[N^5\Bigl(\omega''+\frac{\omega'}{r}+\frac{d}{2-d}\frac{\omega'^2}{\omega}\Bigr)+N^4(\omega'N'-\dot\omega\dot N\Bigr],\label{4-3}
\end{eqnarray}
with the  dot representing now $\partial_\tau$. Note that we are dealing here now with hyperbolic PDE's in stead of elliptic ones.

The exact solution is again given by the solution of Eq.(\ref{A-3}) and Eq.(\ref{A-4}) (in the pseudo-Riemannian spacetime)\footnote{The replacement $t\rightarrow i\tau$ has no influence on the solution.}.
We repeat that our original pseudo-Riemannian 5D spacetime delivers  an effective 4D Riemannian spacetime on the brane 
\begin{equation}
ds^2=\frac{N_1^2}{N_2^2}(dt^{*2}+dr^{*2})+dz^2+r^2d\varphi^{*2},\label{4-4}
\end{equation}
where $r$ can be expressed in $r^*$ (Appendix B). 
The topology is $\mathds{R}^2_+\times\mathds{R}^1\times S^1$. The $\varphi^*$ causes no problems, because the $N^\varphi$ component decouples from the equations for $N$ and $\omega$. $\mathds{R}^2_+$ is still conformally flat.

It is legitimate to conjecture that in the non-vacuum case, the scalar field will also be the same on the Riemannian spacetime. 
This could be used in order to solve the issue of construction of the quantum wave functions as elements of the Fock space.
One of the entangled particles travels on the Klein surface on the hypersurface embedded in the 5D spacetime and remains a pure state. 
In the next sections, we will enlighten this issue.
%======================================================================
\subsection{Some other aspects of the Klein bottle}
%==================================================================
In our model, we need the Klein surface $\mathds{K}$, the compact sum of two projective planes.
Our conjecture is, that our solution can be seen as a dynamical solution embedded in our $(4+1)$ spacetime. More precisely, if ${\cal N}$ is the sub manifold with metric $g$, i.e., the metric  on our effective 4D spacetime, then ${\cal N}$ is diffeomorphic to the hyperbolic 5D spacetime: $\chi :\mathds{R}^5\rightarrow {\cal N}$.
The Riemannian manifold $({\cal N},g)$ must be conformally flat, which is the case here.

Now physicists are interested in the topology of moduli spaces of self-dual connections on vector bundles over Riemannian manifolds. One reason was that on these spaces the instanton approximation to the Green functions of Euclidean quantum gravity Yang-Mills theory can be expressed in terms of integrals over moduli spaces. One needs then the metric and volume form of the moduli spaces.

From the investigations of Groisser, et.al,\cite{gross1987}, we conjecture that we can consider $\mathds{K}$ as a 4-sphere in our hyperbolic pseudo Riemannian spacetime. This suspicion is fueled by the  solution of section 4.2, and the work on the orientable counterpart model of Groisser (and references therein).

More precisely, on a Riemannian 5-manifold, one can proof that there exists a coordinate diffeomorphism $\xi: \mathds{R}^5 \rightarrow {\cal N} $ for which the pullback of the metric $\mathpzc{g}$ on ${\cal N}$ is given by $(\xi^*\mathpzc{g})_{ij}=\Psi^2({\bf x})\delta_{ij}$. Further, the Riemannian manifold $({\cal N},\mathpzc{g})$
is conformally flat with finite radius and  volume. The action of $SO(5)$ on $S^4$ induces an isometry on ${\cal N}$ whose pullback, via $\xi$, is the usual action $SO(5)$ action on $\mathds{R}^5$. So ${\cal N}$ can  isometrically included as the interior of a compact Riemannian manifold with boundary, say $\bar{\cal N}$, whose boundary $\partial \bar{\cal N}$ is isometric to the 4-sphere of constant radius. The embedding $\partial \bar{\cal N}\rightarrow {\cal N}$ is totally geodesic. The sphere $\partial \bar{\cal N}$ is conformally equivalent to the original manifold $(S^4,\mathpzc{g})$ and points on $\partial \bar{\cal N}$ corresponds to instantons which are concentrated at a single point on $S^4$.
It is remarkable that the $\Phi({\bf x})$ is determined by a PDE which is comparable with our scalar equation (${\cal N}$ has no constant curvature).

A remark must be made about the $\mathds{Z}_2$ symmetry in the original description of the antipodal mapping\cite{thooft2021}. 
At the border of region I and II in the Penrose diagram, the antipodes on a 3-sphere were glued together and the transverse  $(\theta,\varphi)$ part is a projected 2-sphere. 
In our model, it is replaced by the projected 3-sphere $(z,\varphi ,\mathpzc{y})$ using the $\mathds{Z}_2$ symmetry of the bulk space $(U,V,z,\varphi,\mathpzc{y})$. No cut and past procedure is necessary.

However, a lot of unsolved issues need to be attacked. In particular, the firewall transformation description in the warped spacetime.
%======================================================================
\section{ Notes on the information paradox}
%========================================================================
There are some known facts about the information paradox.
An outside observer will register the Hawking radiation as thermal, i.e., in a mixed state, while a local observer is in doubt about the vacuum state.
Further, the evolution of the wave function of the in-falling particle must be unitary, i.e., it satisfy the Schr\"odinger equation  $\ket{\chi(t_1)}=U(t_1,t_2)\ket{\chi(t_2)}$ and is bijective. It is believed that during black hole evaporation,  information of the quantum state is preserved. Information loss is inconsistent with unitarity.
So the problem is how to handle the controversy between the pure quantum state of the in-falling particles and the mixed state property of the Hawking radiation.

In our model we do not rely on replica wormholes, but instead on the warped spacetime and Klein surface.

By the antipodal mapping in $\mathds{R}^4$, we have also $e^{in\varphi}\rightarrow e^{in(\varphi +\pi)}$.
This can be "visualized" by considering in figure 3 (right) points on the circle on $S^3$ (for example $\alpha\sim e^{in\varphi}$) for which
\begin{equation}
\sin(\theta/2)e^{in\varphi}{\cal V}(\alpha)-\cos(\theta/2){\cal W}(\alpha)=0.\label{5-1}
\end{equation}
From the real part, we obtain the plane after the stereographic projection, 
\begin{equation}
x_3=\tan(\theta/2)(\cos(\varphi) x_1-\sin(\varphi) x_2.\label{5-2}
\end{equation}
If we apply $e^{in\varphi}\rightarrow e^{in(\varphi +\pi)}$, the plane is rotated over $\pi$.
The imaginary part delivers the two spheres ${\bf x}^2\mp\tan(\theta/2)(\sin(\varphi) x_1+\cos(\varphi)x_2)=1$.

We found in section 3 that the in-falling  Hawking particle travels for a while  on the Klein surface in $\mathds{R}^4$. Now consider the state
\begin{equation}
\ket{\chi}=\ket{\hat\chi} e^{in\varphi}\label{5-3}
\end{equation}
and the density matrix 
\begin{equation}
\rho_A=\sum_{i}p_i\ket{\chi_i}_A{_A}{\bra{\chi_i}}\label{5-4}
\end{equation}
as mixture of pure states and  weights $p_i$  with $\sum p_i=1$. Further,${\bf Tr}\rho_A=1$ and $\rho_A^2=\rho_A$. The density matrix doesn't contain the phase. So the change of the azimuthal angle $\varphi$ has no influence on the pure states. This can be compared with the the "pureness" of $\rho$ in Eq(\ref{3-6}) of normalized state vectors on $S^2\subset \mathds{R}^3$. In addition to $\vec{{\bf V}}$
\begin{equation}
\vec{{\bf V}}=\Bigl({\cal V}{\cal\bar W} +{\cal W}{\cal\bar V} ,i({\cal V}{\cal\bar W} -{\cal W}{\cal\bar V}),\vert{\cal V}\vert^2-\vert{\cal W}\vert^2\Bigr),\label{5-5}
\end{equation}
independent of a phase change in $\ket{{\cal U}}=({\cal V},{\cal W})$, we define
\begin{equation}
\vec{{\bf Z}}=\vec{{\bf P}}+i\vec{{\bf Q}}=\Bigl({\cal V}^2-{\cal W}^2,i({\cal V}^2+{\cal W}^2),-2{\cal V}{\cal W}\Bigr).\label{5-6}
\end{equation}
$(\vec{{\bf V}},\vec{{\bf P}},\vec{{\bf Q}})$ form a positively oriented orthonormal triad. Now $\vec{{\bf P}}$ is a unit tangent vector to $S^2$ at 
$\vec{{\bf V}}$.
If $\ket{{\cal U}}\rightarrow e^{i\alpha}\ket{{\cal U}}$, then $\vec{{\bf V}}$ is independent of this phase change, while  $\vec{{\bf Z}}\rightarrow e^{2i\alpha}\vec{{\bf Z}}$.  One can verify that $\vec{{\bf P}}$ registers phase change $mod (\pi)$ only. When $\ket{{\cal U}}\rightarrow -\ket{{\cal U}}$, $\vec{{\bf U}}$ still delivers the same $\vec{{\bf P}}$, while $(\vec{{\bf R}},\vec{{\bf P}})$ determine $\ket{{\cal U}}$ up to sign.
Let us now interpret  $\vert\braket{{\cal U}}{{\cal U'}}\vert^2$ (independent of the phase) as a "probability". In language of the geometric picture here, 
$\sqrt{1-\vert\braket{{\cal U}}{{\cal U'}}\vert}=\frac{1}{2}\vert\vec{{\bf V}}-\vec{{\cal {\bf V'}}}\vert$ represents the diameter of $S^2$.

This "visualization"is in fact a Hopf fibration of $S^3$ (see also section 3.1)
In our case, we are dealing with a time depended situation and state vectors follow the Schr\"odinger equation
\begin{equation}
\frac{d}{dt}\ket{{\cal U}}(t)=-\frac{1}{\hbar}H\ket{{\cal U}}(t),\label{5-7}
\end{equation}
with $H$ a Hermitian matrix, with $H=H_0\mathds{I}+\vec{{\bf H}}.{\bf \sigma}$.
One finally obtains
\begin{equation}
\frac{d{\vec{{\bf V}}}}{dt}=\vec{{\bf V}}=\frac{2}{\hbar}\vec{{\bf H}}\times \vec{{\bf V}},\label{5-8}
\end{equation}
the time evolution in $S^2$. From Eq.(\ref{5-8}) we have $\vec{{\bf V}}.\vec{{\bf V}}=0$, so pure states remain pure.
%===================================================================
\subsection{Another argument}
%====================================================================
There is another strong argument for considering the topology of the Klein surface.
In short, we blow up the 4-manifold to 5D in order to handle the singularities in the curvature and to apply the antipodal map.
One can mathematically formulate the topology of a 4-manifold using self-dual connections over de Riemannian $S^4$\cite{freed1984}. It depends only upon the conformal class of the Riemannian metric. This self-dual connection  can be interpreted by the conformal map: $\mathds{R}^4\rightarrow S^4/\{0\}$ as a self-dual connection or "instanton".

In our 5D RS model (with a finite number of singularities), we found  the metric is determined by $N$ and $\omega$ (see appendix A). The solution for $N$ on the effective 4D spacetime is the same, while the $\omega$ contribution is different. This is solely due to the contribution of ${\cal E}_{\mu\nu}$ (Eq.(\ref{2-7})).
If we switch to Riemannian $(t\rightarrow i\tau$, a Wick rotation), the solution is for both $N$ and $\omega$ unaltered. Moreover the $\tilde g_{\mu\nu}$ is conformally flat.
The embedding of the Klein surface was done using the extra dimension ${\bf \mathpzc{y}}$ and  the effective spacetime is conformally flat.

Consider now the ball $B^4\subset\mathds{R}^4$ with the induced metric $\omega^4 g_{\mu\nu}$ (Eq.(\ref{A-3}))\footnote{Remember that we used twice the dilaton separation.}. So the scale is $\omega^4$.
Then the ball $\frac{r}{2\omega}B^4\subset\mathds{R}^4$ has a curvature $\leq \frac{const}{\vert\omega\vert^4}$
We interpret the Riemannian 5D warped spacetime Eq.(\ref{4-1}), an open 5-ball,  as an instanton on $\mathds{R}^4$. 

In the pseudo-Riemannian spacetime of Eq.(\ref{2-5}), the boundary is the non-orientable Klein surface, which we used for the antipodicity in order to maintain the pure states of the Hawking particles.

The importance of the instanton trick, is the fact that in the Riemannian space (or easier, in Minkowski space)  they play a crucial role in calculating path integrals\cite{felsager1998}. 
It turns out that a static solution in $m$ space dimensions is completely equivalent to an instanton in $m$ space-time dimensions.
%================================================================
\section{High-frequency approximation close to the horizon}
%============================================================
Our time evolution was governed by the Klein-Gorden-type PDE. 
Although we found a time dependent exact and numerical solution, one should like to describe the quantum effects more precisely.
It is a nice exercise to derive the Schr\"odinger equation from the Klein-Gordon, using the so-called multiple-scale (or high-frequency (HF)) method (see appendix E).

However, we are dealing here also with the polar angle dependency. One usually deals with separability of the wave function in a $(t,r)$ part and spherical harmonics\footnote{They are needed to describe the gravitational back reaction. See for example ref\cite{thooft2019}. }. In our case, this is not possible without extra assumptions.
The HF method can offer a solution for this situation (see, for example, Slagter\cite{slagter2022d}, section 3.3). 
In some sense, one can keep track of the back reaction of the emitted Hawking particles. In the appendix E.4, we applied the method for the Vaidya spacetime in Eddington-Finkelstein coordinates
\begin{equation}
ds^2=-\Bigl(1-\frac{2M(u)}{r}\Bigr)du^2-2dudr+r^2(d\theta^2+\sin^2\theta d\varphi^2),\label{5-9}
\end{equation}
which is the Schwarzschild spacetime with $u=t-r-2M\log(r/2M-1)$. This spacetime is used to demonstrate the loss of mass by gravitational radiation.
From Eq.(\ref{E-49}) and  Eq.(\ref{E-50}) we obtain
\begin{equation}
h_{rr}=h_{r\theta}=h_{r\varphi}=0,\qquad h_{\varphi\varphi}=-\sin^2\theta h_{\theta\theta}\label{5-10}
\end{equation}
\begin{eqnarray}
\ddot k_{rr}=0,\qquad \dot h_{\theta\theta}=r\partial_r\dot h_{\theta\theta},\qquad
\dot h_{\theta\varphi}=r\partial_r\dot h_{\theta\varphi}.\label{5-11}
\end{eqnarray}
So one writes
\begin{equation}
h_{\theta\theta}=r\alpha(u,\theta,\varphi,\xi),\quad h_{\theta\varphi}=r\beta(u,\theta,\varphi,\xi), \quad h_{\varphi\varphi}=-r\alpha\sin^2\theta.\label{5-12}
\end{equation}
Further, we have
\begin{eqnarray}
\ddot k_{r\theta}=\frac{1}{r}\Bigl(2\dot\alpha\cot\theta +\partial_\theta\dot\alpha+\frac{1}{\sin^2\theta}\partial_\varphi\dot\beta\Bigr),\label{5-13}
\end{eqnarray}
\begin{eqnarray}
\ddot k_{r\varphi}=\frac{1}{r}\Bigl(\dot\beta\cot\theta -\partial_\varphi\dot\alpha +\partial_\theta\dot\beta\Bigr),\label{5-14}
\end{eqnarray}
\begin{eqnarray}
\frac{dM}{du}=-\frac{\ddot k_{\phi\phi}+\sin^2\theta\ddot k_{\theta\theta}}{4\sin^2\theta}
-\frac{1}{2}r\dot h_{uu}-\frac{1}{4}\Bigl(\dot\alpha^2+\frac{\dot\beta^2}{\sin^2\theta}\Bigr)
+\frac{1}{4}\Bigl(\ddot{\alpha^2}+\frac{\ddot{\beta^2}}{\sin^2\theta}\Bigr).\label{5-15}
\end{eqnarray}
Not all the components of $h_{\mu\nu}$ and $k_{\mu\nu}$ are physical, so one needs some extra gauge conditions.
Suitable choice of $\alpha$ and $\beta$ (Choquet-Bruhat uses, for example, $ \alpha=0,  \beta=g(u)h(\xi)\sin\theta$), leads to a solution to second order which is in general not axially symmetric. We can integrate these zero order equations with respect to $\xi$. One obtains then some conditions on the background fields, because  terms like $\int \dot\alpha d\xi$ disappear.
From Eq.(\ref{5-15}), we obtain
\begin{equation}
\frac{dM}{du}=-\frac{1}{4\tau}\int_0^\tau\Bigl(\dot\alpha^2+\frac{\dot\beta^2}{\sin^2\theta}\Bigr)d\xi,\label{5-16}
\end{equation}
which is the back-reaction of the high-frequency disturbances on the mass $M$. $\tau$ is the period of $\dot h_{\mu\nu}$. This expression can be substituted back into Eq.(\ref{5-15}).
However, in the non-vacuum case, the right-hand side will also contain contributions from the matter fields.
In order to obtain propagation equations for $h_{\mu\nu}$ and $k_{\mu\nu}$, one  proceeds with the next order equation $R_{\mu\nu}^{(1)}=0$. First of all, Eq.(\ref{5-13}), (\ref{5-14}) are consistent with $R_{r\varphi}^{(1)}=0$ and $R_{r\theta}^{(1)}=0$.
Further, one obtains  propagation equations for $\alpha$ and $\beta$ and for some second order perturbations, such as $k_{\varphi\varphi}$. Moreover, the $(\varphi, \theta)$-dependent part of the PDE's for $\alpha$ and $\beta$
(say $A(\theta,\varphi), B(\theta,\varphi)$) can be separated (for the case $k_{\theta\varphi}\neq 0$): 
\begin{equation}
\partial_\varphi B+2\sin\theta\cos\theta A+\sin^2\theta\partial_\theta A=0,\label{5-17}
\end{equation}
\begin{eqnarray}
\sin^2\theta\partial_{\theta\theta}A+7\sin\theta\cos\theta\partial_\theta A+4\cot\theta\partial_\varphi B +2(5\cos^2\theta-1)A+2\partial_{\theta\varphi}B=0.\label{5-18}
\end{eqnarray}
A non-trivial simple solution is
\begin{equation}
A=\frac{\cos\theta (sin\varphi+\cos\varphi)}{sin^3\theta},\quad B=\frac{\sin\varphi-\cos\varphi}{\sin^2\theta}+G(\theta),\label{5-19}
\end{equation}
with $G(\theta)$ arbitrary. So the breaking of the spherically and axial symmetry is evident.
%=========================================================================
\section{Conclusion}\label{sec5}
%===========================================================================================
On a warped 5D Randall-Sundrum Kerr-like black hole spacetime, we investigated the time evolution of the scalar-gauge field in the conformal invariant gravity model. We analyzed the dilaton-Higgs $(\omega, \Phi)$ system and compared the numerical solution with the exact vacuum counterpart model found in an earlier study. It turns out that the presentation of the model on a Klein surface, fits very well in the antipodal interpretation of the black hole paradoxes.
We find that the potential $V(\omega,\Phi)$ is determined by the superfluous dilaton equations. A connection is made with the instanton solution on the effective 4D Riemannian space.
There remain a lot of questions. 
First, could one explain that fluctuations in the dilaton generate the spacetime as noticed  by the outside observer?
Secondly, the zero's of the quitic polynomial $N_1(r)$, which determine the singular point of the spacetime, can be written as (in the complex plane) without a scalar field
\begin{equation}
\frac{c+10r^2(r-a)^3-10r^3(r-a)^2+r^4(r-a)-r^5}{5r^2}
\end{equation}
When $r\rightarrow a$, we are left with a singularity at $r=0$ when $a$ becomes very small  in the far future $t\rightarrow t_H$ (Eq.(\ref{A-4})). Could $a$ be related to the mass of the black hole in its end-time? When the scalar gauge field is included, one should like to find an exact solution, guided by the vacuum exact solution and  the numerical investigation.
Thirdly, one can apply an approximation scheme in order to obtain the graviton exchange on the horizon. This is intensive work. 
It could be possible that the eikonal approximation must be replaced by the two-timing method, or high-frequency approximation.
Fourth, the construction of the Fock space in the warped spacetime must be investigated more deeply. 
It is conjectured that the $\mathds{Z}_2$ symmetry of the bulk space is necessary in order to construct a consistent description of evaporation of the black hole. 
Further, the role of the dilaton in connection with the information paradox, must be investigated. This "quantum"field $\omega$  determines the vacuum state of the local observer.
These issues are currently under investigation by the author.
%===================================================================================
\backmatter
\bmhead{Acknowledgments}
This research was solely done under ASFYON, which is financial independent. We thanks  several fellow researchers for valuable comments.
=======================oOo==================
%===============================================================================================
%===========================================================================================
\begin{appendices}
%=========================================================================================
\section{Former results without a scalar field}\label{secA1}
%===========================================================================
In a former study\cite{slagter2018,slagter2021,slagter2022c} we found in the case without a scalar field, that an exact solution exists for $\omega$ and $N$.
The equation for the dilaton was obtained from the Einstein equations and was of the elliptic type (d=4,5):
\begin{equation}
\ddot\omega=-N^4\omega''+\frac{d}{\omega(d-2)}\Bigl(N^4\omega'^2+\dot\omega^2\bigr).\label{A-1}
\end{equation}
The equation for $N$ is
\begin{eqnarray}
\ddot N=\frac{3\dot N^2}{N}-N^4\Bigl(N''+\frac{3N'}{r}+\frac{N'^2}{N}\Bigr)\hspace{4.7cm}\cr
-\frac{d-1}{(d-3)\omega}\Bigl[N^5\Bigl(\omega''+\frac{\omega'}{r}+\frac{d}{2-d}\frac{{\omega'}^2}{\omega}\Bigl)+N^4\omega' N'+\dot\omega\dot N\Bigr].\label{A-2}
\end{eqnarray}
The exact solution is
\begin{equation}
\omega=\Bigl(\frac{a_1}{(r+a_2)t+a_3r+a_2a_3}\Bigr)^{\frac{1}{2}d-1},\label{A-3}
\end{equation}
\begin{eqnarray}
N^2\equiv N_1(r)^2 N_2(t)^2 \qquad\qquad\qquad\qquad\qquad \cr =\frac{10a_2^3r^2+20a_2^2r^3+15a_2r^4+4r^5+C_1}{5C_2r^2(a_3+t)^4+C_3}=\frac{4\int r(r+a_2)^3dr}{r^2[C_2(a_3+t)^4+C_3]},\label{A-4}
\end{eqnarray}
with $a_i,C_i$ some constants. $a_2$ is related to the mass.
Note that the solution for $N$ is the same in the bulk and on the brane.
The singularities of the spactime are determined by a quintic. For more details of the zeros and the remarkable relation with Klein surface, we refer to Slagter\cite{slagter2022c}.

It is remarkable that one can write the solution for $N_1(r)$ (see second expression in Eq.(\ref{A-4})) as a result of a first-order differential equation;
\begin{equation}
N_1^2+rN_1\frac{\partial N_1}{\partial r}=2(r+a_2)^3.\label{A-5}
\end{equation}
Without the bulk contribution, it becomes
\begin{equation}
N_1^2+rN_1\frac{\partial N_1}{\partial r}=\frac{3}{2}(r+a_2)^2.\label{A-5a}
\end{equation}
We also compared the solution with the Ba\v nados-Teitelboim-Zanelli (BTZ) black hole solution in $(2+1)$-dimensional spacetime (without the $dz^2$ term).
The solution could be written as
\begin{equation}
N_1^2=\frac{4}{l^2r^2}\int r(r+l\sqrt{4GM})(r-l\sqrt{4GM}),\label{A-5b}
\end{equation}
solution of the first order differential equation
\begin{equation}
N_1^2+rN_1\frac{\partial N_1}{\partial r}=\frac{1}{2l^2}(r+l\sqrt{4GM})(r-l\sqrt{4GM}).\label{A-5c}
\end{equation}
Here $l$ is the scale at which curvature sets in and is related to the cosmological constant $\Lambda=-\frac{1}{l^2}$.
It seems that the solution on the brane, without the bulk contribution (Eq.(\ref{A-5a})), must be related to the BTZ solution of Eq.(\ref{A-5c}).
%===================================================================================
\section{Kruskal coordinates and the Penrose diagram}
%===================================================================================
We define the coordinates, 
\begin{equation}
dr^*\equiv\frac{1}{N_1(r)^2}dr\quad  dt^*\equiv N_2(t)^2dt.\label{A-6}
\end{equation}
Our induced effective spacetime can be written as\cite{slagter2022c}
\begin{equation}
ds^2=\omega^{4/3}\bar\omega^2\Bigl[\frac{N_1^2}{N_2^2}\Bigl(-dt{^*}^2+dr{^*}^2  \Bigr) +dz^2+r^2(d\varphi+\frac{N^\varphi}{N_2^2} dt^*)^2 \Bigr],\label{A-6}
\end{equation}
due to the fact that the solution for the metric component $N$ is a quotient $\frac{N_1(r)}{N_2(t)}$.
Further,
\begin{eqnarray}
r^*=\frac{1}{4}\sum_{r^H_i}\frac{r^H_i \log(r-r^H_i)}{(r^H_i+b_2)^3},\qquad t^*=\frac{1}{4C_2}\sum_{t^H_i}\frac{\log(t-t^H_i)}{(t^H_i+b_3)^3}\label{A-7}.
\end{eqnarray}
The sum it taken over the roots of $(10b_2^3r^2+20b_2^2r^3+15b_2r^4+4r^5+C_1)$ and $C_2(t+b_3)^4+C_3$, i. e., $r^H_i$ and $t^H_i$.

This polynomial in $r$ defining the roots of $N_1^2$, is a  quintic equation, which has some interesting connection with Klein's  icosahedron solution. 
Further, one can define the azimuthal angular coordinate $d\varphi^*\equiv (d\varphi +\frac{N^\varphi}{N_2^2} dt^*)$, which can be used when an incoming null geodesic falls into the event horizon. $\varphi^*$  is the azimuthal angle in a coordinate  system rotating about the z-axis relative to the Boyer-Lindquist coordinates.
Next, we define the coordinates \cite{strauss2020} (in the case of $C_1=C_3=0$ and 1 horizon, for the time being) 
\begin{eqnarray}
U_+=e^{\kappa (r^*-t^*)}, \hspace{1.2cm} V_+=e^{\kappa (r^*+t^*)} \hspace{1cm} r>r_H \cr
U_-=-e^{\kappa (r^*-t^*)}, \hspace{0.85cm} V_-=-e^{\kappa (r^*+t^*)}\hspace{0.6cm} r<r_H,\label{A-8}
\end{eqnarray}
with $\kappa$ a constant.
The spacetime becomes
\begin{equation}
ds^2=\omega^{4/3}\bar\omega^2\Bigl[\frac{N_1^2}{N_2^2}\frac{dUdV}{\kappa^2UV}+dz^2+r^2 d\varphi^{*2}\Bigr].\label{A-9}
\end{equation}
The antipodal points $P(X)$ and $ P(\bar X)$ are physically identified. If we compactify the coordinates,
\begin{equation}
\tilde U=\tanh U,\qquad \tilde V=\tanh V,\label{A-10}
\end{equation}
then the spacetime can  be written as
\begin{equation}
ds^2=\omega^{4/3}\bar\omega^2\Bigl[H(\tilde U,\tilde V)d\tilde U d\tilde V +dz^2+r^2 d\varphi^{*2}\Bigr],\label{A-11}
\end{equation}
with
\begin{equation}
H=\frac{N_1^2}{N_2^2}\frac{1}{\kappa^2 \arctanh \tilde U \arctanh\tilde V(1-\tilde U^2)(1-\tilde V^2)}.\label{A-12} 
\end{equation}
We can write $r$ and $t$ as
\begin{equation}
r=r_H+\Bigl(\arctanh \tilde U \arctanh \tilde V\Bigl)^{\frac{1}{2\kappa\alpha}},\quad
t=t_H+\Bigl(\frac{\arctanh \tilde V}{ \arctanh \tilde U}\Bigl)^{\frac{1}{2\kappa\beta}},\label{A-13}
\end{equation}
with 
\begin{equation}
\alpha=\frac{r_H}{4(r_H+b_2)^3}, \qquad \beta=\frac{1}{4C_2(t_H+b_3)^3}.\label{A-14}
\end{equation}
Further, $N_1$ and $N_2$ can be expressed in $( U, V)$ (or $(\tilde U,\tilde V)$). One easily checks that for $U=0$ and $V=0$ one  obtains the horizons as depicted in the Penrose diagram of figure B1.
Now $N_1$ and $N_2$ are polynomials in $r$ and $t$. So for $U=0$ and $V=0$ they approach $r_H$ and $t_H$ as expected. So the "scale-term" $H$ is consistent with the features of the Penrose diagram.
%=========================================================================================
\begin{figure}[h]
\centerline{
\includegraphics[width=1.\textwidth]{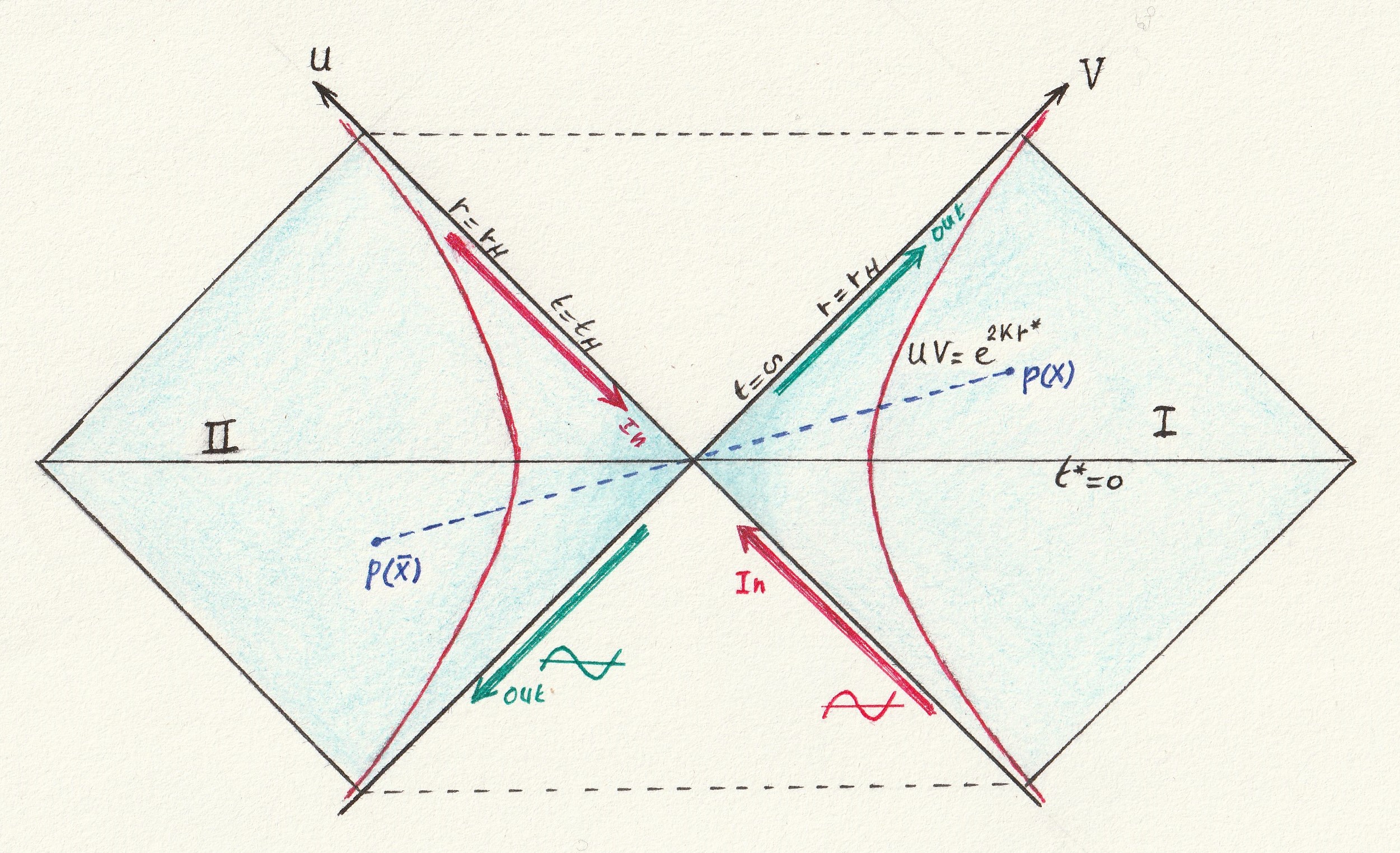}}
\caption{Penrose diagram. The antipodal points $P(X)$ and $P(\bar X)$ are identified. Particles going in will generate waves  approaching the horizon from the outside and passes  the horizon, will reappear  from  "the other side" of the black hole. Note that $t^*=log(U/V)\sim log(t-t_H)$.}
\end{figure}
Note that $ds^2$ and $H$ are invariant under $\tilde U\rightarrow -\tilde U$ and $\tilde V\rightarrow -\tilde V$.
$\tilde g_{\mu\nu}$ is regular everywhere and conformally flat.
Note that the hyperbolas in the Penrose diagram (see Figure B1)  can be described by
\begin{equation}
UV=e^{2\kappa r^*}=(r-r_H)^{\frac{\kappa r_H}{2(r_H+b_2)^3}},\label{A-15}
\end{equation}
where $b_2$ contains  the mass and $\kappa$ a constant. So when the black hole is evaporating, the hyperbola moves to the left  in region I in the Penrose diagram. If $r$ is close to $r_H$, the hyperbola approaches the Planckian area.

%===========================================================================================
\section{ In $(U,V)$-coordinates}
%====================================================================================
In light cone coordinates $u=r-t$ and $v=r+t$, one obtains a set of PDE's
\begin{eqnarray}
\Delta N=\frac{1}{1+N^4} \mathpzc{A},\quad \Delta\omega =\frac{1}{1-N^4}\mathpzc{B},\quad \Delta X=\frac{1}{1-N^4} \mathpzc{C},\label{A-16}
\end{eqnarray}
where $\mathpzc{A}, \mathpzc{B}, \mathpzc{C}$ are expressions in $N, \omega ,X$ and their derivatives. Further, $\Delta\equiv \partial_u^2+\partial_v^2$.
These  PDE's can be used, in a numerical setting, in order to follow the Hawking particles in the $(U,V)$ plane.
\footnote{This technical aspect is currently under investigation by the author.}
One also can proof that Laplace equations make sense on Klein surfaces\cite{alling1971}
%==========================================================================
\section{Notations}\label{secA2}
%==========================================================================
$d$: spacetime  dimension\\
$n$:  topological windingnumber or Chern number \\
$(t,r,z,\varphi)$: coordinates of the 4D spacetime\\
$\mathpzc{y}$: fifth dimension in RS model\\
$y_0$: dimension of the bulk spacetime\\
$\Bigl(N(t,r), N^\varphi(t,r)\Bigr)$: metric components\\
${\cal V}$: $z+i\mathpzc{y}$\\
${\cal W}$: $x+iy$\\
${\cal Z}: \frac{{\cal V}}{{\cal W}}$\\
$\mathds{R}^n$: n-dimensional Euclidean space\\
$\mathds{C}^n$: complex  n-dimensional space\\
$\eta$: vacuum expectation value\\
$\omega(t,r)$: dilaton field\\
$\Phi=\eta X(t,r) e^{in\varphi}$: complex scalar (Higgs) field\\
$A_\mu$: gauge field =$(0,0,0,A_\varphi,0)$\\
$S^m$: m-sphere, $\{ x\in \mathds{R}^{m+1}; \vert x\vert =1\}$\\
$B^m$: m-ball\\
$\mathds{R}P^m$: real projective m-space\\
$\mathds{C}P^m$: complex projected m-space\\
$\mathds{Z}_2$: reflection symmetry, i.e. a symmetry operation that yields the identity if applied twice\\
$\mathds{T}$: torus $S^1\times S^1$\\
$\mathds{K}:$ Klein surface = compact surface  (i.e., triangulation possible) of the connected sum of two projective planes $\mathds{R}P^1\# \mathds{R}P^1$\\
$f:S^2\rightarrow \mathds{R}P^1$: map of 2-sphere onto its quotient space (or identification space topology)\\
$(S^2,f)$: covering space of $\mathds{R}P^1$
%==============================================================================================
\section{The Schr\" odinger equation and the multiple-scale method}
%============================================================================================
\subsection{ The $\Phi^4$-Klein-Gordon equation}
%===================================================================================
Consider the 1+1 dimensional Lagrangian density\footnote{We follow the notation of Felsager\cite{felsager1998}. Here one can find more details of the short presentation here.}
\footnotesize{
\begin{equation}
{\cal L}=\frac{1}{2}\Bigl[-\partial_\mu\Phi\partial^\mu\Phi-V(\Phi\Bigr],\label{E-0}
\end{equation}}
with $V=\frac{1}{4}\beta(\Phi^2-\frac{\alpha}{\beta})^2 $.
From Euler-Lagrange equations we obtain the well-known $\Phi^4$ model 
\begin{equation}
\Phi_{tt}-\Phi_{xx}=-(\beta\Phi^2-\alpha)\Phi.\label{E-1}
\end{equation}
Note that $-\frac{1}{2}\alpha\Phi^2$ is the mass term in the potential.
The potential possesses two distinct minima, so we have a classical degenerated vacuum, $\Phi_0=\pm \sqrt{\frac{\alpha}{\beta}}$. At infinity, the field is static and the configuration must have finite energy. This results in boundary conditions

\begin{eqnarray}
\Phi_{\pm\infty}=\lim_{x\rightarrow\pm\infty}\Phi(x,t)=\pm \sqrt{\frac{\alpha}{\beta}},\qquad U(\Phi_{\pm\infty})=0.\label{E-2}
\end{eqnarray}
%=========================================================================================
\subsection{The "kink"-sector}
%===========================================================================================
The most simple topological soliton solution of the $\Phi^4$ model occur in one space dimension.
Solutions interpolate between the two vacua $\Phi_{\pm}=\sqrt{\frac{\alpha}{\beta}}$. The topological "charge" is then
\begin{equation}
Q=\int_{-\infty}^{+\infty}\frac{\partial\Phi}{\partial x}dx=\Phi(+\infty,t)-\Phi(-\infty,t)=\frac{\Phi_+ -\Phi_-}{2\sqrt{\frac{\alpha}{\beta}}},\quad Q=\{-1,0,1\}.\label{E-2a}
\end{equation}
We see that $Q$ only depends on the asymptotic behavior at infinity. 
Note that in the related {\it sine-Gordon} model, one has an infinite series of discrete minima, (Q = 0...n) in contrast with the $\Phi^4$ model. So in this case Q is topologically {\it quantized}. However, in 2 space dimensions $(r,\varphi)$ in the axially symmetry, one has the topological quantization via the {\it winding number}\footnote{Note that solitary waves doesn't exists in 2 space dimensions (Derrick's theorem). So one needs the vortex solution (Nielsen-Oleson string)}.

One can wonder, if the vacuum degeneracy will survive the quantization. The answer is yes. The soliton solution is quantum-mechanically stable, i.e., cannot decay into one of the vacua.
%===================================================================
\begin{figure}[h]
\centerline{
\includegraphics[width=8cm]{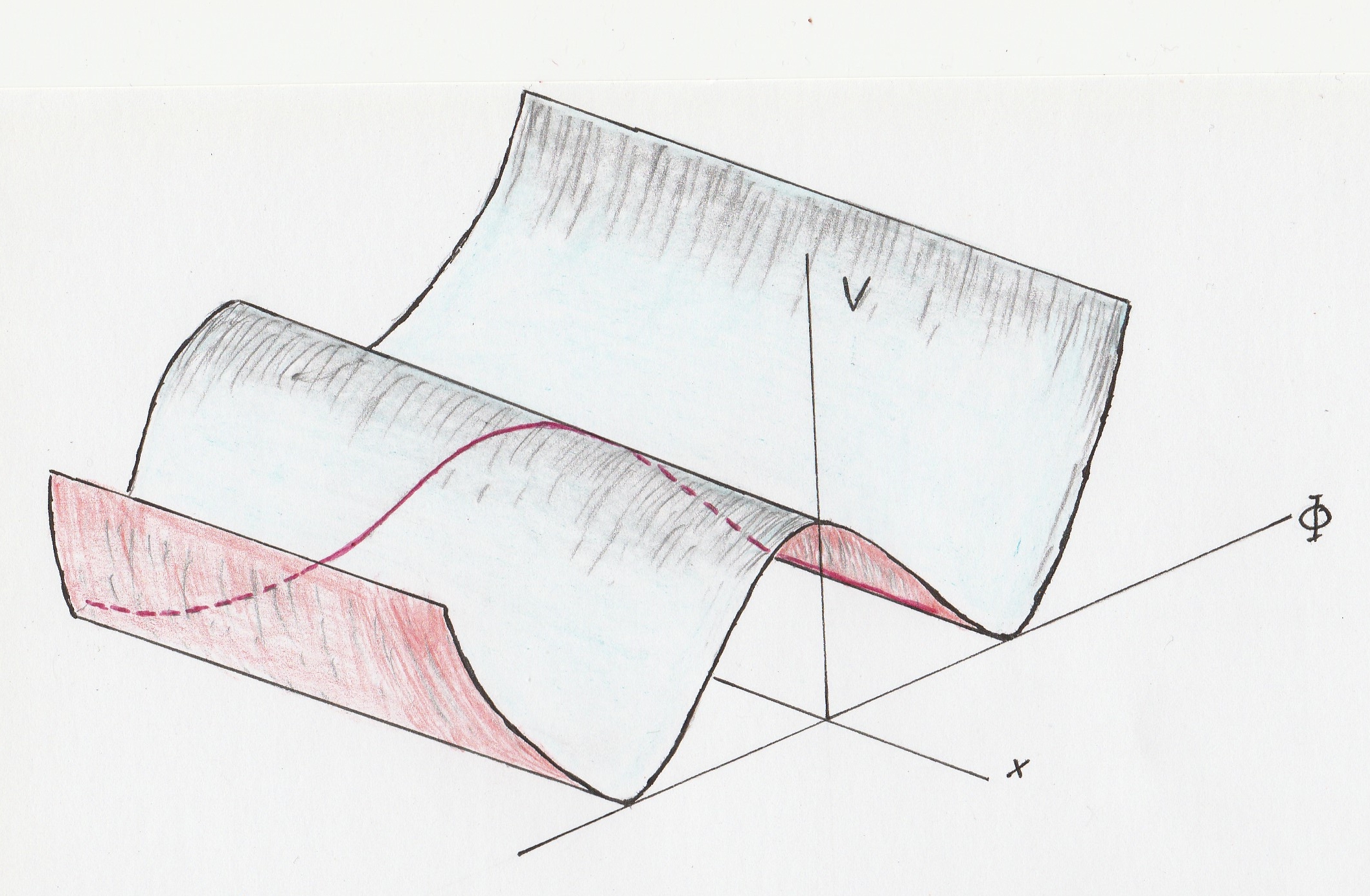}}
\caption{The $\Phi$-kink needs an infinite amount of energy in order to overcome the ridge of infinite length.}
\end{figure}
%======================================================================
A {\it traveling wave} solution can easily  be obtained.

If one wants a localized  traveling wave solution of the form (solitary wave) 
\begin{equation}
\Phi(x,t)=f(x-vt)=f(\xi),\qquad \lim_{\xi\rightarrow \pm\infty }f(\xi)=\Phi_{\pm\infty},\label{E-3}
\end{equation}
then one can write the equation as
\begin{equation}
\frac{df}{d\xi}=\pm\gamma\sqrt{2U[f]},\qquad \gamma=\frac{1}{\sqrt{1-v^2}},\label{E-4}
\end{equation}
or
\begin{equation}
\lambda(\xi-\xi_0)=\pm\int\frac{df}{\sqrt{2U[f]}}.\label{E-5}
\end{equation}
It possesses the "kink" and "anti-kink"-solutions.
%========================================================================================
\subsection{The method of multiple-scales}
%====================================================================================
Let us expand
\begin{equation}
\Phi(x,t;\epsilon)=\sum_{1}^{\infty}\epsilon^{kp}\Phi^{(k)}(x,t;\epsilon)\label{E-6}
\end{equation}
with $\epsilon$ a small parameter.  Introduce "slow" variables
\begin{equation}
{\rm X}_k=\epsilon^k x, \qquad {\rm T}_k=\epsilon^k t,\quad k=1,2,.\label{E-7}.
\end{equation}
So we write
\begin{equation}
\Phi(x,t;\epsilon)=\tilde\Phi(x,t,{\rm X}_1 ,{\rm T}_1,{\rm X}_2,{\rm T}_2,...;\epsilon)\label{E-8}
\end{equation}
and
\begin{equation}
\frac{\partial\Phi}{\partial x}=\frac{\partial\tilde\Phi}{\partial x}
+\sum_{k=1}\epsilon^k\frac{\partial\tilde\Phi}{\partial {\rm X}_k},\qquad
 \frac{\partial\Phi}{\partial t}=\frac{\partial\tilde\Phi}{\partial t}
+\sum_{k=1}\epsilon^k\frac{\partial\tilde\Phi}{\partial {\rm T}_k}\label{E-9}
\end{equation}
So $\epsilon$ measures  the ratio of the fast scale to the slow scale.
Substitution into Eq.(\ref{E-1}) with also
\begin{equation}
\tilde\Phi(x,t,{\rm X}_1,{\rm T}_1,...;\epsilon)
 =\sum_{k=1}^{\infty}\epsilon^{kp}\tilde\Phi^{(k)}(x,t,{\rm X}_1,{\rm T}_1,...;\epsilon)\label{E-10}
\end{equation}
and equating coefficients of equal powers of $\epsilon$, results in:(tilde  omitted):

\underline {1. term of ${\cal O}(\epsilon^p)$}
\begin{equation}
\epsilon^p\Bigl(\frac{\partial^2}{\partial x^2}-\frac{\partial^2}{\partial t^2}
+\alpha\Bigr)\Phi^{(1)}\label{E-11}
\end{equation}
\underline {2. term of ${\cal O}(\epsilon^{p+1})$}
\begin{equation}
2\epsilon^{p+1}\Bigl(\frac{\partial^2}{\partial x\partial{\rm X}_1}
-\frac{\partial^2}{\partial t\partial{\rm T}_1}\Bigr)\Phi^{(1)}\label{E-12}
\end{equation}
\underline {3. term of ${\cal O}(\epsilon^{p+2})$}
\begin{equation}
\epsilon^{p+2}\Bigl(\frac{\partial^2}{\partial {\rm X}_1^2}-\frac{\partial^2}{\partial {\rm T}_1^2}
+2\frac{\partial^2}{\partial x\partial{\rm X}_2}-
2\frac{\partial^2}{\partial t\partial{\rm T}_2}\Bigr)\Phi^{(1)}\label{E-13}
\end{equation}
\underline {4. term of ${\cal O}(\epsilon^{2p})$}
\begin{equation}
\epsilon^{2p}\Bigl(\frac{\partial^2}{\partial x^2}-\frac{\partial^2}{\partial t^2}
+\alpha\Bigr)\Phi^{(2)}\label{E-14}
\end{equation}
\underline {5. term of ${\cal O}(\epsilon^{2p+1})$}
\begin{equation}
2\epsilon^{2p+1}\Bigl(\frac{\partial^2}{\partial x\partial{\rm X}_1}
-\frac{\partial^2}{\partial t\partial{\rm T}_1}\Bigr)\Phi^{(2)}\label{E-15}
\end{equation}
\underline {6. term of ${\cal O}(\epsilon^{2p+2})$}
\begin{equation}
\epsilon^{2p+2}\Bigl(\frac{\partial^2}{\partial {\rm X}_1^2}-\frac{\partial^2}{\partial {\rm T}_1^2}
+2\frac{\partial^2}{\partial x\partial{\rm X}_2}-
2\frac{\partial^2}{\partial t\partial{\rm T}_2}\Bigr)\Phi^{(2)}\label{E-16}
\end{equation}
\underline {7. term of ${\cal O}(\epsilon^{3p})$}
\begin{equation}
\epsilon^{3p}\Bigl((\frac{\partial^2}{\partial x^2}-\frac{\partial^2}{\partial t^2}
+\alpha\Bigr)\Phi^{(3)} +\epsilon^{3p}\beta(\Phi^{(1)})^3\label{E-17}
\end{equation}
So to lowest order we get:
\begin{equation}\boxed{
\Bigl(\frac{\partial^2}{\partial x^2}-\frac{\partial^2}{\partial t^2}
+\alpha\Bigr)\Phi^{(1)}=0}\label{E-18}
\end{equation}
This is the {\it linearized equation}.
The next term is ${\cal O}(\epsilon^{p+1})$ and/or $ {\cal O}(\epsilon^{2p})$. However
$2p<(p+1)$ is impossible, because $p\in{\rm N}$. So for the next term we have  $p+1<2p$
or $p+1=2p$. If $p+1=2p$ then $p=1$ and we have (we ignore for the moment the other possibility $p+1<2p, i.e., p=2$ or 3 or 4, ...)
for the next order:
\begin{equation}
\Bigl(\frac{\partial^2}{\partial x^2}-\frac{\partial^2}{\partial t^2}+\alpha\Bigr)\Phi^{(2)}+2
\Bigl(\frac{\partial^2}{\partial x\partial{\rm X}_1}
-\frac{\partial^2}{\partial t\partial{\rm T}_1}\Bigr)\Phi^{(1)}=0\label{E-19}
\end{equation}
and\footnotesize{
\begin{eqnarray}
\Bigl(\frac{\partial^2}{\partial x^2}-\frac{\partial^2}{\partial t^2}+\alpha\Bigr)\Phi^{(3)} +
\Bigl(\frac{\partial^2}{\partial {\rm X}_1^2}-\frac{\partial^2}{\partial {\rm T}_1^2}
+2\frac{\partial^2}{\partial x\partial{\rm X}_2} \cr
-2\frac{\partial^2}{\partial t\partial{\rm T}_2}\Bigr)\Phi^{(1)} -\beta(\Phi^{(1)})^3
+2 \Bigl(\frac{\partial^2}{\partial x\partial{\rm X}_1}
-\frac{\partial^2}{\partial t\partial{\rm T}_1}\Bigr)\Phi^{(2)}=0\label{E-20}
\end{eqnarray}}
One can solve in principle Eq.(\ref{E-18})-Eq.(\ref{E-20}). From  Eq.(\ref{E-18})
we have the solution
\begin{equation}
\Phi^{(1)}(x,t,{\rm X}_1,{\rm T}_1,{\rm X}_2,{\rm T}_2,...)=
A^{(1)}({\rm X}_1,{\rm T}_1,{\rm X}_2,{\rm T}_2,...)e^{i\hbar(kx-\nu t)} + CC\label{E-21}
\end{equation}
with $\hbar^2(k^2-\nu^2) =\alpha$ ({\it dispersion relation}), $k$ the wave number (number of oscillations per $2\pi$ in space) and $\nu$ the frequency (number of oscillations per $2\pi$ in time). Substitution in Eq.(\ref{E-19}) yields
\begin{equation}\boxed{
\Bigl(\frac{\partial^2}{\partial x^2}-\frac{\partial^2}{\partial t^2}+\alpha\Bigr)\Phi^{(2)} =
-2i\hbar \Bigl(k\frac{\partial A^{(1)}}{\partial{\rm X}_1}
+\nu\frac{\partial A^{(1)}}{\partial{\rm T}_1}\Bigr)e^{i\hbar(kx-\nu t)} +CC}\label{E-22}
\end{equation}
The solution will contain a {\it secular} term of the form
\begin{equation}
B({\rm X}_1,{\rm T}_1,...)(kx-\nu t)e^{i\hbar(kx-\nu t)}\label{E-23}
\end{equation}
due to the fact of resonance with $e^{i\hbar(k'x-\nu ' t)}$,
which disturbs the asymptotic behavior. So we take
\begin{equation}
\Bigl(k\frac{\partial A^{(1)}}{\partial{\rm X}_1}
+\nu\frac{\partial A^{(1)}}{\partial{\rm T}_1}\Bigr)=0\label{E-24}
\end{equation}
That means, $A^{(1)}$ will be of the form
\begin{equation}
A^{(1)}({\rm X}_1,{\rm T}_1,{\rm X}_2,{\rm T}_2,...)=A^{(1)}(\xi,{\rm X}_2,{\rm T}_2,...),\label{E-25}
\end{equation}
with
\begin{equation}
\xi\equiv{\rm X}_1-\frac{k}{\nu}{\rm T}_1.\label{E-26}
\end{equation}
So we have as first {\it approximate solution} of Eq.(\ref{E-1})
\begin{eqnarray}
\Phi^{(1)}(x,t;\epsilon )&=&A^{(1)}\Bigl({\rm X}_1-\frac{k}{\nu}
{\rm T}_1,{\rm X}_2,{\rm T}_2,...\Bigr)
e^{i\hbar(kx-\omega t)}+CC \cr
&=&A^{(1)}\Bigl(\epsilon(x-\frac{k}{\nu} t),\epsilon^2x,\epsilon^2 t,...\Bigr)
e^{i\hbar(kx-\omega t)}+CC \cr
&\approx& A^{(1)}\Bigl(\epsilon(x-\frac{k}{\nu} t),0,0,...\Bigr)e^{i\hbar(kx-\nu t)}+CC.\label{E-27}
\end{eqnarray}
We recognize the well-known carrier-wave (amplitude modulation)-solution: the amplitude varies
slowly with respect to the wave $e^{i\hbar(kx-\nu t)}$. The envelope $A^{(1)}$ has a velocity of $\frac{k}{\nu}=\frac{d\nu}{dk}$, i. e., the velocity of the wave-packet.

Now we want to find the expression for $A^{(1)}$. From Eq.(\ref{E-22}) and Eq.(\ref{E-24}) we obtain
\begin{equation}
\Bigl(\frac{\partial^2}{\partial x^2}-\frac{\partial^2}{\partial t^2}+\alpha\Bigr)\Phi^{(2)}=0,\label{E-29}
\end{equation}
so a solution
\begin{equation}
\Phi^{(2)}=A^{(2)}({\rm X}_1,{\rm T}_1,{\rm X}_2,{\rm T}_2,...)e^{i\hbar(k'x-\nu ' t)} + CC\label{E-30}
\end{equation}
with $\hbar^2(k'^2-\nu'^2)=\alpha$.
Substitution in Eq.(\ref{E-20}) yields:\footnotesize{
\begin{eqnarray}
\Bigl(\frac{\partial ^2}{\partial x^2}-\frac{\partial ^2}{\partial t^2}+\alpha\Bigr)\Phi^{(3)}
=-\Bigl(1-\frac{k^2}{\nu ^2}\Bigr)\frac{\partial^2A^{(1)}}{\partial\xi^2}e^{i\hbar(kx-\omega t)}\cr
-2i\hbar\Bigl(k\frac{\partial A^{(1)}}{\partial {\rm X}_2}+\nu\frac{ \partial A^{(1)}}
{\partial{\rm T}_2}\Bigr)e^{i\hbar(kx-\nu t)} - 2i\hbar\Bigl(k'\frac{\partial A^{(2)}}
{\partial{\rm X}_1}+\nu '\frac{\partial A^{(2)}}{\partial{\rm T}_1}\Bigr)e^{i\hbar(k'x-\nu 't)}\cr
+\beta\Bigl((A^{(1)})^3 e^{3i(kx-\nu t)}+3(A^{(1)})^2\bar A^{(1)}e^{i(kx-\nu t)}\Bigr) +CC\qquad\qquad\label{E-31}
\end{eqnarray}}
Again we skip the {\it secular terms}:

\begin{eqnarray}
1.\quad k'\frac{\partial A^{(2)}}{\partial {\rm X}_1}+\nu '\frac{\partial A^{(2)}}{\partial{\rm T}_1}=0\label{E-32}
\end{eqnarray}
so we have
\begin{equation}
A^{(2)}({\rm X}_1,{\rm T}_1,{\rm X}_2,{\rm T}_2,...)=A^{(2)}(\xi '.{\rm X}_2,{\rm T}_2,...),\label{E-33}
\end{equation}
with $\xi '={\rm X}_1-\frac{k'}{\nu '}{\rm T}_1$,
and
\begin{equation}
2.\quad \Bigl(1-\frac{k^2}{\nu ^2}\Bigr)\frac{\partial^2A^{(1)}}{\partial\xi ^2}
+2i\hbar\Bigl(k\frac{\partial A^{(1)}}{\partial{\rm X}_2}+\nu\frac{\partial A^{(1)}}{\partial{\rm T}_2}\Bigr)
-3\beta\vert A^{(1)}\vert^2A^{(1)}=0.\label{E-34}
\end{equation}
If we introduce the moving coordinate system (group-velocity $\frac{k}{\omega}$) $\xi_2={X}_2-\frac{k}{\nu} { T}_2$  and $T_2$, in order to get rid of $\partial X_2$, we obtain for the amplitude
\begin{equation}
A^{(1)}(\xi  ,\xi_2,{\rm T}_2,{\rm X}_3,{\rm T}_3,...)\approx A^{(1)}(\epsilon(x-\frac{k}{\nu}t),
\epsilon^2(x-\frac{k}{\nu}t),\epsilon^2 t,...)\label{E-35}
\end{equation}
the partial differential equation:
\begin{equation}\boxed{
i\hbar \frac{\partial A^{(1)}}{\partial{\rm T}_2}-\frac{\alpha}{2\hbar^2 \nu^3}\frac{\partial^2 A{(1)}}
{\partial \xi_2 ^2}-\frac{3\beta}{2\nu}\vert A^{(1)}\vert^2A^{(1)} =0}.\label{E-36}
\end{equation}
which represents a {\it cubic Schr\" odinger equation}. We used the dispersion relation $k^2-\omega^2=\frac{\alpha}{\hbar^2}$.
A typical solution  is obtained for $\alpha=-2\hbar^2\omega^3$
\begin{equation}
A^{(1)}=C_1 Exp\Bigl[i\Bigl(C_2\xi_2-(\frac{3\beta}{2\omega}C_1^2+C_2^2)T_2+C_3\Bigr)\Bigr]\label{E-37}
\end{equation}
where $C_i$ are constants. 
So  we have the  first order approximation  solution
for $\Phi^{(1)}$
\begin{equation}\boxed{
\Phi^{(1)}(x,t;\epsilon)=\Phi^{(1)}(x,t,\xi,\xi_2,T_2,...)=C_1 e^{i(C_2\xi_2-(\frac{3\beta}{2\omega}C_1^2+C_2^2)T_2+C_3)}e^{i\hbar(k x-\omega t)}+CC}\label{E-38}
\end{equation}
This form can be compared with the so-called "amplitude modulation".
Now in our model, $\alpha$ is related to the mass in this $\Phi^4$ model. 
Other interesting solutions are
\begin{equation}
A^{(1)}=2C_1\sqrt{\frac{\omega}{3\beta}}\frac{e^{i(C_1^2T_2+C_2)}}{\cosh(C_1\xi_2+C_3)}\label{E-39}
\end{equation}
\begin{equation}
A^{(1)}=2C_1\sqrt{\frac{\omega}{3\beta}}\frac{e^{i\Bigl(C_2\xi_2+(C_1^2-C_2^2)T_2+C_1\Bigr)}}{\cosh(C_1\xi_2-2C_1C_2T_2+C_3)}\label{E-40}
\end{equation}
\begin{equation}
A^{(1)}=\frac{C_1}{\sqrt{T_2}}e^{i\Bigl[\frac{(\xi_2+C_2)^2}{4T_2} +(\frac{3\beta}{2\omega}C_1^2\ln T_2+C_3) \Bigr]}\label{E-41}
\end{equation}
We have now two scales, the slow $(\xi_2, T_2)$  and the fast $(t, x)$.
Most of the solutions are obtained by the inverse-scattering method, the method of Zakharov-Shabat.
Further, there are conserved  quantities,
\begin{eqnarray}
\int_{-\infty}^{+\infty}\vert\Phi\vert^2 dx,\quad \int_{-\infty}^{+\infty} i(\bar \Phi\partial_x\Phi-\Phi\bar{\partial\Phi})dx,\quad \int_{-\infty}^{+\infty}\Bigl(\vert\partial_x\Phi\vert^2+\frac{3\beta}{4\omega} \vert\Phi\vert^4\Bigr)\label{E-43}
\end{eqnarray}
It would be interesting to explore the case $p=2$.
%=================================================================
\subsection{Application to the  Vaidya spacetime}
%============================================================
One expands the relevant fields
\begin{equation}
V_i=\sum_{n=0}^{\infty}\frac{1}{\nu^n}F_i^{(n)}({\bf x},\xi),\label{E-44}
\end{equation}
where $\nu$ represents a dimensionless  parameter ("frequency"), which will be large. Further, $\xi=\nu\Theta({\bf x})$, with $\Theta$ a scalar (phase) function on the manifold.
The small parameter $\frac{1}{\nu}$ can also be the ratio of the characteristic wavelength of the perturbation to the characteristic dimension of the background. On warped spacetimes it could also be the ratio of the extra dimension to the background dimension.
In the vacuum case, we expand the metric
\begin{equation}
g_{\mu\nu}=\bar g_{\mu\nu}+\frac{1}{\nu}h_{\mu\nu}({\bf x},\xi)+\frac{1}{\nu^2}k_{\mu\nu}({\bf x},\xi)+ ...,\label{E-45}
\end{equation}
where we defined
\begin{eqnarray}
\frac{dg_{\mu\nu}}{dx^\sigma}=g_{\mu\nu,\sigma}+\nu l_\sigma\dot g_{\mu\nu},\quad
g_{\mu\nu,\sigma}=\frac{\partial g_{\mu\nu}}{\partial x^\sigma},\quad 
\dot g_{\mu\nu}=\frac{\partial g_{\mu\nu}}{\partial \xi},\label{E-46}
\end{eqnarray}
with $l_\mu =\frac{\partial\Theta}{\partial x^\mu}$.
One then says that
\begin{equation}
V_i=\sum_{n=-m}^{\infty}\frac{1}{\nu^n}F_i^{(n)}({\bf x},\xi)\label{E-47}
\end{equation}
is  an approximate wavelike solution of order n of the field equation, if $F_i^{(n)}=0, \forall n$. One can substitute the expansion into the field equations.
The Ricci tensor then expands as
\begin{equation}
R_{\mu\nu}\rightarrow \omega R_{\mu\nu}^{(-1)}+\Bigl(\bar R_{\mu\nu}+R_{\mu\nu}^{(0)}\Bigr)+\frac{1}{\omega}R_{\mu\nu}^{(1)} + ...\label{E-48}
\end{equation}
By equating the subsequent orders to zero, we obtain
\begin{equation}
R_{\mu\nu}^{(-1)}=0=\frac{1}{2}\bar g^{\beta\lambda}(l_\lambda l_\mu\ddot h_{\beta\nu}+l_\nu l_\beta\ddot h_{\mu\lambda}-l_\lambda l_\beta\ddot h_{\mu\nu}-l_\nu l_\mu\ddot h_{\beta\lambda}),\label{E-49}
\end{equation}
\begin{eqnarray}
R_{\mu\nu}^{(0)}+\bar R_{\mu\nu}=0,\qquad  R_{\mu\nu}^{(1)}=0,\qquad ....\label{E-50}
\end{eqnarray}
The bar stands for the background.
Here we used $l_\mu l^\mu =0$. The rapid variation is observed  in the direction of $l_\mu$. 
The eikonal condition, $l_\mu l^\mu =0$, in linear approximations, is adopted as gauge condition. In the high-frequency approximation, however,  it follows from the highest order ($n=-1$) equations.
In the radiative outgoing Eddington-Finkelstein coordinates, we have $x^1=u=\Theta({\bf x})=t-r$ and $ l_\mu =(1,0,0,0)$.\\
=========================oOo==========================
%========================================================================
\end{appendices}
%======================================================================

%=========================================================================
\end{document}